\documentclass[bj]{imsart}

\usepackage{verbatim,graphicx,amssymb,amsmath}
\usepackage{natbib}

\usepackage{graphicx}
\usepackage[lined,algoruled,noline]{algorithm2e}
\startlocaldefs
\newtheorem{theorem}{Theorem}

\newtheorem{corollary}[theorem]{Corollary}

\newtheorem{lemma}[theorem]{Lemma}

\newenvironment{proof}[1][Proof]{\noindent\textbf{#1.} }{\ \rule{0.5em}{0.5em}}

\newcommand{\ba}{\begin{array}}
\newcommand{\ea}{\end{array}}
\newcommand{\be}{\begin{eqnarray}}
\newcommand{\ee}{\end{eqnarray}}
\newcommand{\bes}{\begin{eqnarray*}}
\newcommand{\ees}{\end{eqnarray*}}
\newcommand{\f}{\mathbf}

\newcommand{\x}{\mathbf{x}}
\newcommand{\y}{\mathbf{y}}
\newcommand{\e}{\mathbf{e}}
\newcommand{\A}{\mathbf{A}}
\newcommand{\M}{\mathbf{M}}
\newcommand{\N}{\mathbf{N}}
\newcommand{\bP}{\mathbf{P}}
\newcommand{\Q}{\mathbf{Q}}
\newcommand{\F}{\mathbf{F}}

\newcommand{\D}{\mathbf{D}}
\newcommand{\G}{\mathbf{G}}
\newcommand{\K}{\mathbf{K}}

\newcommand{\bmu}{{\boldsymbol\mu}}
\newcommand{\bnu}{{\boldsymbol\nu}}
\newcommand{\btheta}{{\boldsymbol\theta}}
\newcommand{\bSigma}{{\boldsymbol\Sigma}}

\newcommand{\E}{\operatorname{E}}
\newcommand{\Var}{\operatorname{Var}}
\newcommand{\Norm}{\operatorname{N}}

\newcommand{\zapthis}[1]{} 
\endlocaldefs

\begin{document}

\begin{frontmatter}

\title{Accelerated Gibbs sampling of normal distributions using matrix splittings and polynomials}

\runtitle{Polynomial accelerated Gibbs sampling}

\begin{aug}
\author{\fnms{Colin} \snm{Fox}\thanksref{a}\corref{}\ead[label=e1]{fox@physics.otago.ac.nz}}
\and
\author{\fnms{Albert} \snm{Parker}\thanksref{b}\ead[label=e2]{parker@math.montana.edu}}
\address[a]{Department of Physics, University of Otago, Dunedin, New Zealand \printead{e1}}
\address[b]{Center for Biofilm Engineering, Department of Mathematical Sciences, Montana State University, Bozeman, MT, USA \printead{e2}}

\runauthor{Fox and Parker}


\end{aug}

\begin{abstract}
Standard Gibbs sampling applied to a multivariate normal distribution with a specified precision matrix is equivalent in fundamental ways to the Gauss-Seidel iterative solution of linear equations in the precision matrix.
Specifically, the iteration operators, the conditions under which convergence occurs, and geometric convergence factors (and rates) are identical. These results hold for arbitrary matrix splittings from classical iterative methods in numerical linear algebra giving easy access to mature results in that field, including existing convergence results for antithetic-variable Gibbs sampling, REGS sampling, and generalizations.  Hence, efficient deterministic stationary relaxation schemes lead to efficient generalizations of Gibbs sampling. The technique of polynomial acceleration that significantly improves the convergence rate of an iterative solver derived from a \emph{symmetric} matrix splitting may be applied to accelerate the equivalent generalized Gibbs sampler. Identicality of error polynomials guarantees convergence of the inhomogeneous Markov chain, while equality of convergence factors ensures that the optimal solver leads to the optimal sampler. Numerical examples are presented, including a Chebyshev accelerated SSOR Gibbs sampler applied to a stylized demonstration of low-level Bayesian image reconstruction in a large 3-dimensional linear inverse problem.
\end{abstract}




\end{frontmatter}

\section{Introduction}
The Metropolis-Hastings algorithm for MCMC was introduced to main-stream statistics around 1990~\citep{RobertCasella2011}, though prior to that the Gibbs sampler provided a coherent approach to investigating distributions with
Markov random field 
structure~\citep{Turcin71,Grenander83,Geman,GelfandSmith,BesagGreen93,Sokal93}.
\zapthis{Many early algorithmic developments came from statistical physics~\citep{BesagGreen93},
including the Metropolis algorithm.  Indeed, \cite{Geman} referred to the Gibbs sampler as a \emph{heat bath} Metropolis, because in statistical physics Gibbs sampling corresponds to \emph{local heat bath thermalization}.}  The Gibbs sampler may be thought of as a particular Metropolis-Hastings algorithm that uses the conditional distributions as proposal distributions, with acceptance probability always equal to 1~\citep{Geyer2011}.

In statistics the Gibbs sampler is popular because of ease of implementation \citep[see, e.g.,][]{RobertsSahu}, when conditional distributions are available in the sense that samples may be drawn from the full conditionals. However, the Gibbs sampler is not often presented as an efficient algorithm, particularly for massive models. In this work we show that generalized and accelerated Gibbs samplers are contenders for the fastest sampling algorithms for normal target distributions, because they are equivalent to the fastest algorithms for solving systems of linear equations.

Almost all current MCMC algorithms, including Gibbs samplers, simulate a \emph{fixed} transition kernel that induces a \emph{homogeneous} Markov chain that converges \emph{geometrically} in distribution to the desired target distribution. In this aspect, modern variants of the Metropolis-Hastings algorithm are unchanged from the Metropolis algorithm as first implemented in the 1950's. The adaptive Metropolis algorithm of \cite{Haarioetal01} \citep[see also][]{RobertsRosenthal2007} is an exception,
though it converges to a geometrically convergent Metropolis-Hastings algorithm that bounds convergence behaviour.

We focus on the application of Gibbs sampling to drawing samples from a multivariate normal distribution with a given covariance or precision matrix. Our concern is to develop generalized Gibbs samplers with optimal geometric, or better than geometric, distributional convergence by drawing on ideas in numerical computation, particularly the mature field of computational linear algebra. We apply the matrix-splitting formalism to show that fixed-scan Gibbs sampling from a multivariate normal is equivalent in fundamental ways to the stationary linear iterative solvers applied to systems of equations in the precision matrix.

Stationary iterative solvers are now considered to be very slow precisely because of their geometric rate of convergence, and are no longer used for large systems. However, they remain a basic building block in the most efficient linear solvers.
By establishing equivalence of error polynomials we provide a route whereby acceleration techniques from numerical linear algebra may be applied to Gibbs sampling from normal distributions. The fastest solvers employ non-stationary iterations, hence the equivalent generalized Gibbs sampler induces an inhomogeneous Markov chain. Explicit calculation of the error polynomial guarantees convergence, while control of the error polynomial gives optimal performance.

The adoption of the matrix splitting formalism gives the following practical benefits in the context of fixed-scan Gibbs sampling from  normal targets:

\begin{enumerate}
	\item a one-to-one equivalence between generalized Gibbs samplers and classical linear iterative solvers;
	\item rates of convergence and error polynomials for the Markov chain induced by a generalized Gibbs sampler;
	\item acceleration of the Gibbs sampler to induce an inhomogeneous Markov chain that achieves the optimal error polynomial, and hence has optimal convergence rate;
	\item numerical estimates of convergence rate of the Gibbs sampler in a single chain and \emph{a priori} estimates of number of iterations to convergence;
	\item access to preconditioning, whereby the sampling problem is transformed into an equivalent problem for which the accelerated Gibbs sampler has improved convergence rate.
\end{enumerate}

Some \emph{direct} linear solvers have already been adapted to sampling from multivariate normal distributions,  with \cite{Rue2} demonstrating the use of solvers based on Cholesky factorization to allow computationally efficient sampling. This paper extends the connection to the \emph{iterative} linear solvers. Since iterative methods are the most efficient for massive linear systems, the associated samplers will be the most efficient for very high dimensional normal targets.

\subsection{Context and overview of results}

The Cholesky factorization is the conventional way to produce
samples from a moderately sized multivariate normal distribution \citep{Rue2,RueHeld},
and is also
the preferred method for solving moderately sized linear systems.  For
\emph{large} linear systems,
iterative solvers are the methods of choice due to their inexpensive cost 
per iteration, and small computer memory requirements. 

Gibbs samplers applied to normal distributions are essentially identical to stationary iterative methods from numerical linear algebra.
This connection was exploited by \cite{Adler}, and independently by \cite{Barone}, who noted that the component-wise Gibbs sampler is a stochastic version of the Gauss-Seidel linear solver, and accelerated the Gibbs sampler by introducing a relaxation parameter to implement the stochastic version of the successive over-relaxation (SOR) of Gauss-Seidel.
This pairing was further analyzed by~\cite{GoodmanSokalMGMC}.

\begin{figure}
\centerline{
\begin{tabular}{cc}
{\bf Solving} $\A\x=\f{b}$ & {\bf Sampling from} $\Norm(\bmu,\A^{-1})$\\
A: Gauss-Seidel & B: Gibbs\\
\includegraphics[width=0.45\textwidth,height=0.3\textwidth]{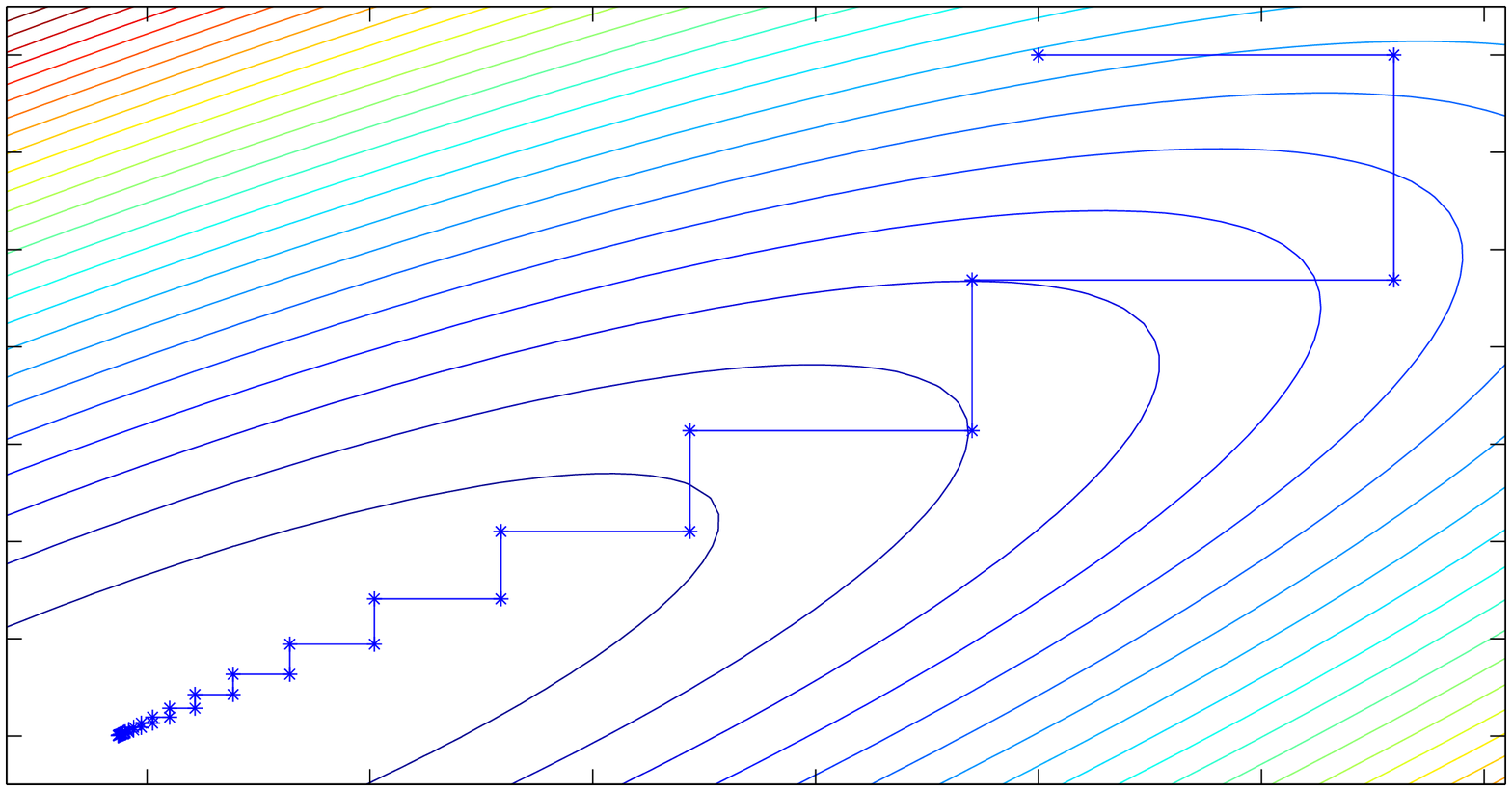} &
\includegraphics[width=0.45\textwidth,height=0.3\textwidth]{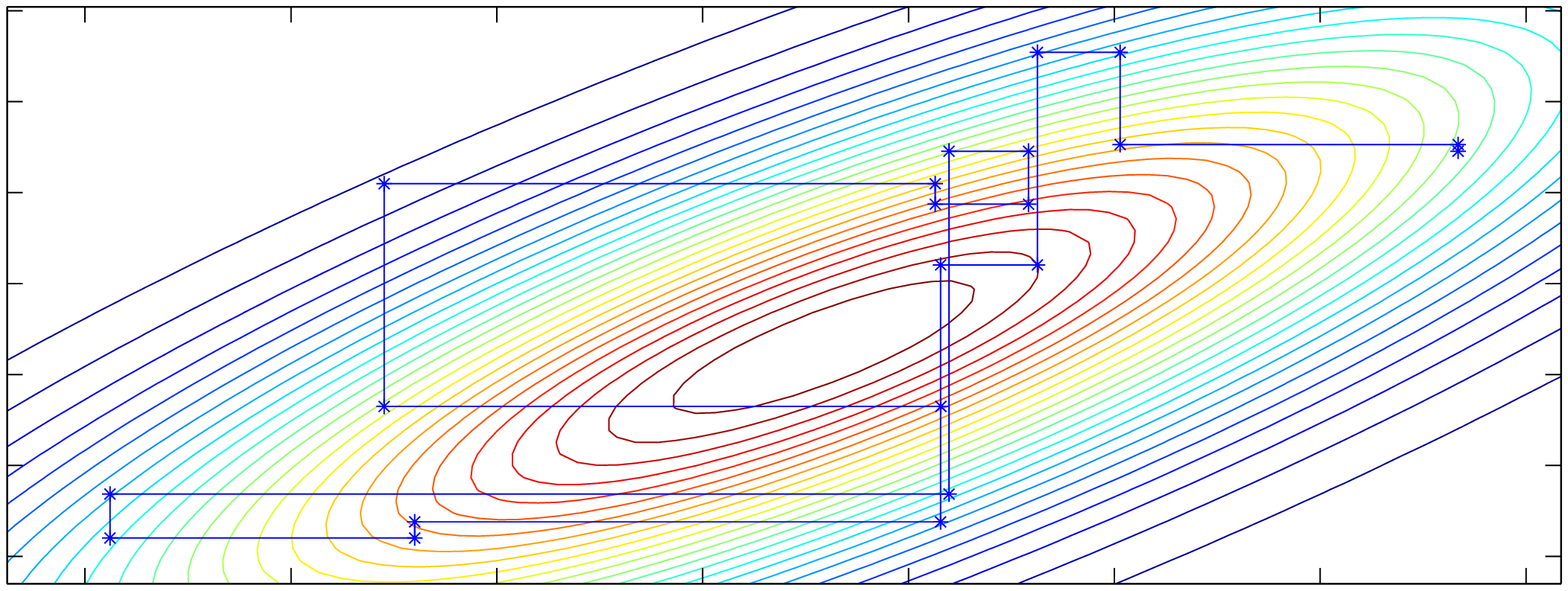}\\
C: Chebyshev-SSOR & D: Chebyshev-SSOR sampler\\
\includegraphics[width=0.45\textwidth,height=0.3\textwidth]{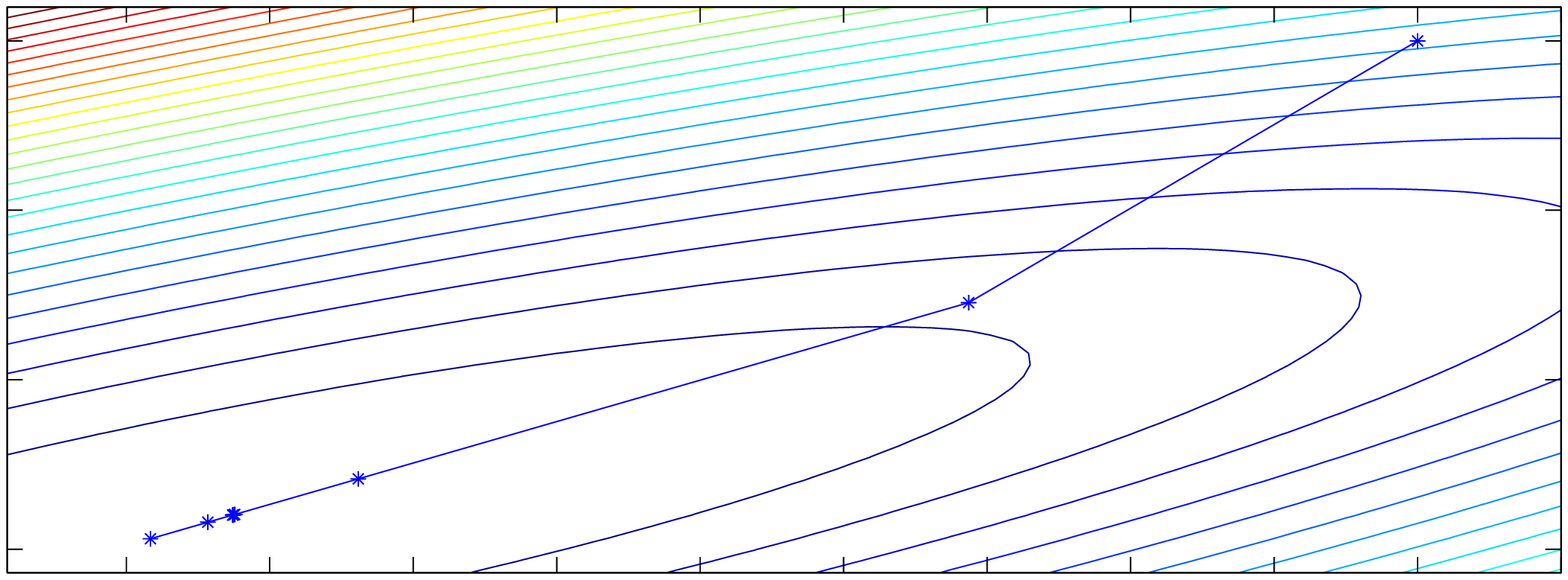} &
\includegraphics[width=0.45\textwidth,height=0.3\textwidth]{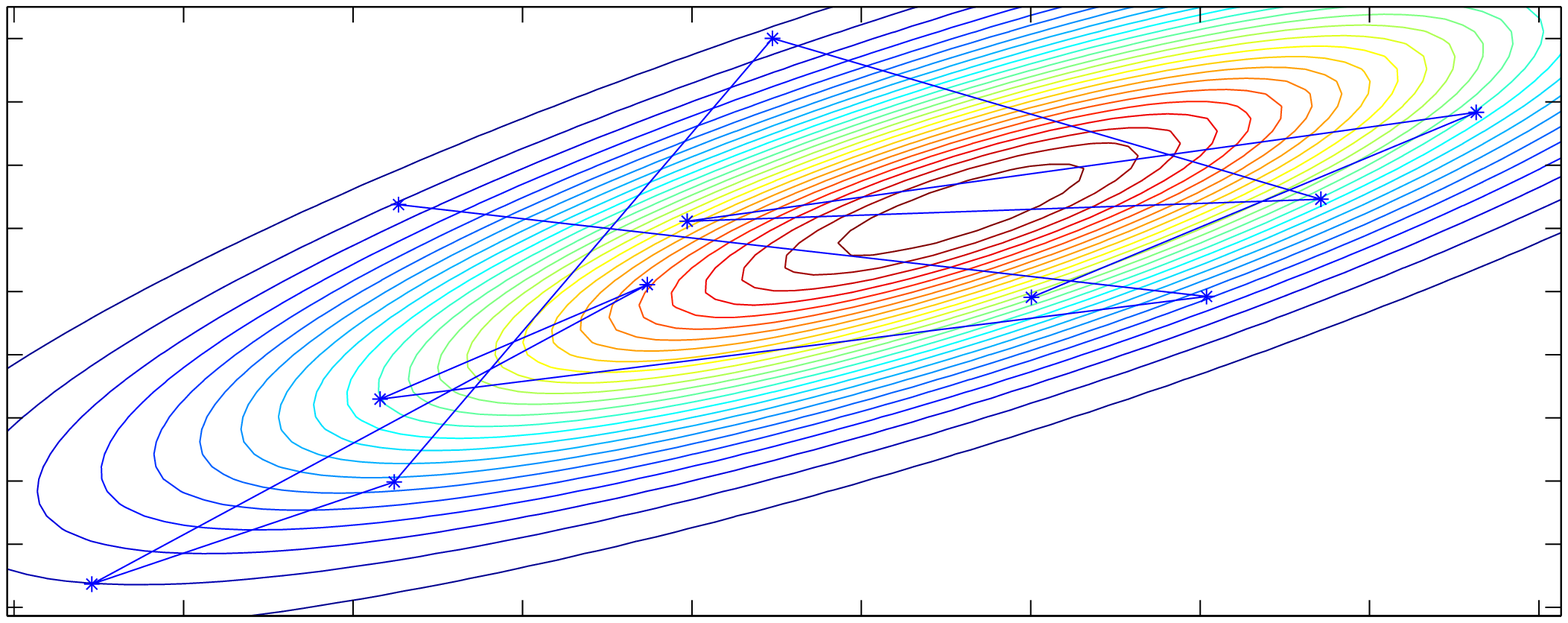}\\
E: CG & F: CG Gibbs\\
\includegraphics[width=0.45\textwidth,height=0.3\textwidth]{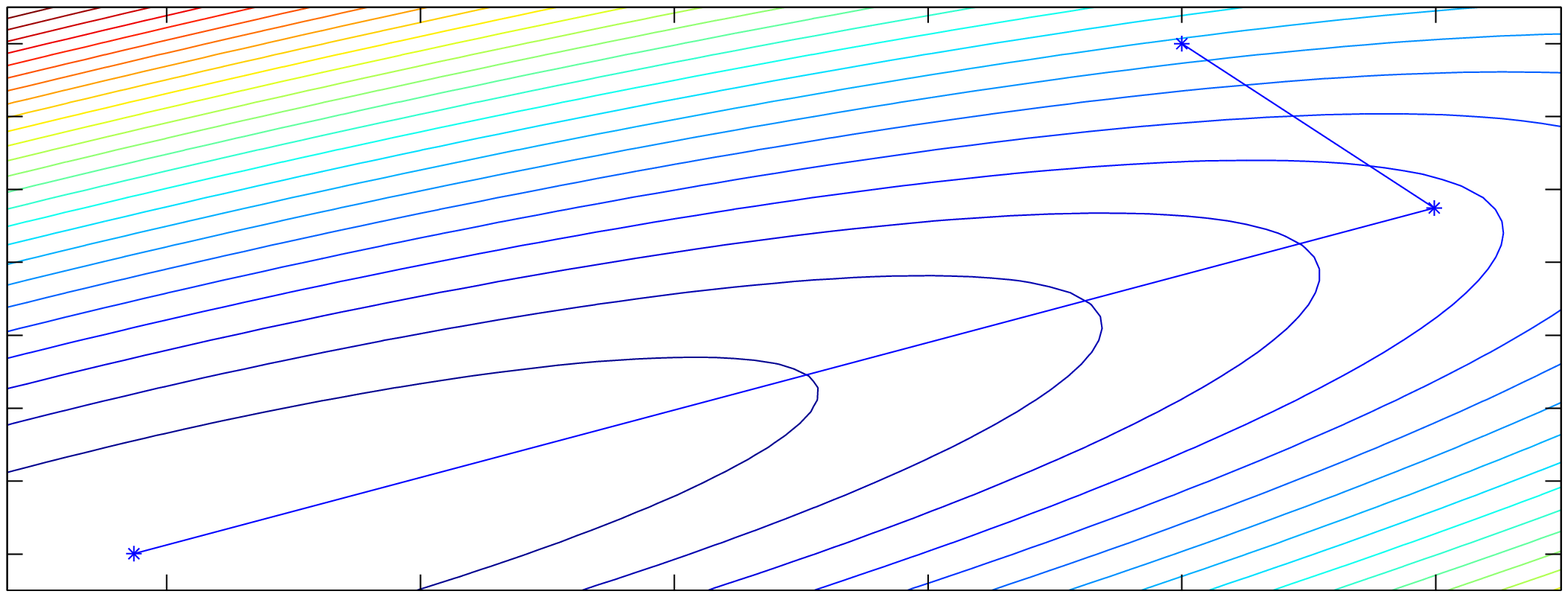} &
\includegraphics[width=0.45\textwidth,height=0.3\textwidth]{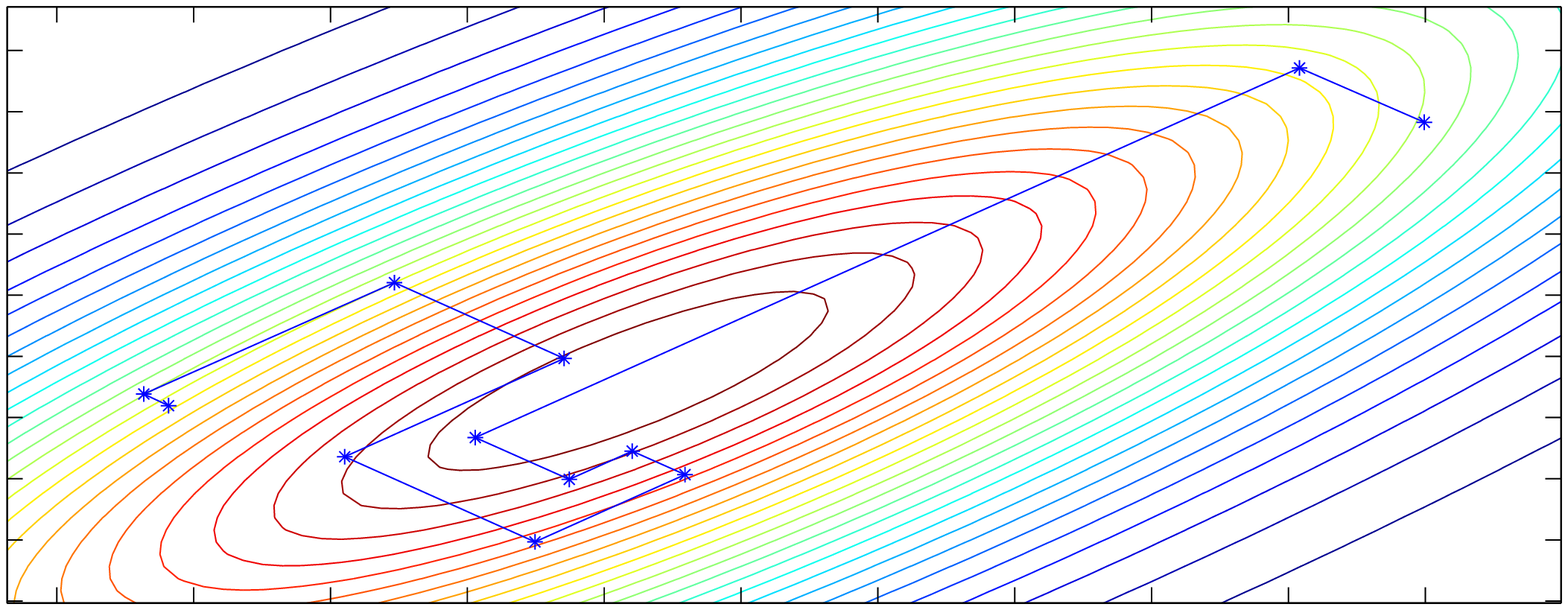}\\
\end{tabular}}
\caption{The panels in the left column show the contours of a quadratic function $\frac{1}{2}\x^T\A\x - \mathbf{b}^T\x$
in two dimensions and the iteration paths for some common optimizers towards the minimizer $\mathbf{\bmu}=\A^{-1}\mathbf{b}$,
or equivalently the path of iterative linear solvers of $\A\x=\mathbf{b}$. The right column presents the iteration paths of the samplers
corresponding to each linear solver, along with the contours of the normal density $k\exp\left\{-\frac{1}{2}\x^T\A\x + \mathbf{b}^T\x\right\}$, where $k$ is the normalizing constant.  In all panels, the matrix $\A$ has eigenpairs $\{\left(10,[1~1]^T\right), \left(1,[1~-1]^T\right)\}$.
The Gauss-Seidel solver took 45 iterations to converge to $\bmu$ (shown are the 90 coordinate steps; each iteration is a ``sweep" of the two coordinate directions), the Chebyshev polynomial accelerated SSOR required just 16 iterations to converge, while CG finds the minimizer in 2 iterations.
For each of the samplers 10 iterations are shown (the 20 coordinate steps are shown for the Gibbs sampler).
The correspondence between these linear solvers/optimizers and samplers is treated in the text (CG in supplementary material).}
  \label{fig:quadraticANDGaussians}
\end{figure}

This equivalence is depicted in panels A and B of Figure~\ref{fig:quadraticANDGaussians}. Panel B shows the contours of a normal density $\pi(\x)$, and  a sequence of coordinate-wise conditional samples taken by the Gibbs sampler applied to $\pi$. Panel A shows the contours of the quadratic minus $\log\left(\pi(\x)\right)$ and the Gauss-Seidel sequence of coordinate optimizations\footnote{Gauss-Seidel optimization was rediscovered by \cite{Besag1986} as iterated conditional modes.}, or, equivalently, solves of the normal equations $\nabla \log\pi(\x)=0$. Note how in Gauss-Seidel the step sizes decrease towards convergence, which is a tell-tale sign that convergence (in value) is geometric.
In Section~\ref{sect:equiv} we will show that the iteration operator is identical to that of the Gibbs sampler in panel B, and hence the Gibbs sampler also converges geometrically (in distribution). Slow convergence of these algorithms is usually understood in terms of the same intuition; high correlations correspond to long narrow contours, and lead to small steps in coordinate directions and many iterations being required to move appreciably along the long axis of the target function.

\cite{RobertsSahu} considered forward then backward sweeps of coordinate-wise Gibbs sampling, with relaxation parameter, to give a sampler they termed the \emph{REGS} sampler.
This is a stochastic version of the symmetric-SOR (SSOR) iteration, which comprises forward then backward sweeps of SOR.

The equality of iteration operators and error polynomials, for these pairs of fixed-scan Gibbs samplers and iterative solvers, allows existing convergence results in
numerical analysis texts~\citep[for example][]{Ax,Golub,Nevanlinna,SaadIter,Young} to be used to establish convergence results for the corresponding Gibbs sampler.
Existing results for rates of distributional convergence by fixed-sweep Gibbs samplers \citep{Adler,Barone,LiuWongKong95,RobertsSahu} may be established this way.


The methods of Gauss-Seidel, SOR, and SSOR, give stationary linear iterations that were used as linear
solvers in the 1950's, and are now considered very slow. The corresponding fixed-scan Gibbs samplers are slow for precisely the same reason. The last fifty years has seen an  explosion of theoretical results
and algorithmic development that have made linear solvers faster and more
efficient, so that for large problems, stationary methods are used
as preconditioners at best, while the method of preconditioned
conjugate gradients, GMRES, multigrid, or fast-multipole methods
are the current state-of-the-art for solving linear systems in a
finite number of steps \citep{Saad20th}.

Linear iterations derived from a symmetric splitting may be sped up by polynomial acceleration, particularly Chebyshev acceleration
that results in optimal error reduction amongst methods that have a fixed non-stationary iteration structure \citep{FoxParker1968,Ax}.
The Chebyshev accelerated SSOR solver and corresponding Chebyshev accelerated SSOR sampler \citep{FoxParkerSISC2014} are depicted in panels C and D of Figure~\ref{fig:quadraticANDGaussians}.
Both the solver and sampler take steps that are more aligned with the long axis of the target, compared to the coordinate-wise algorithms, and hence achieve faster convergence.
However, the step size of Chebyshev-SSOR solving still decreases towards convergence, and hence convergence for both solver and sampler is still asymptotically geometric, albeit with much improved rate.

\cite{FoxParkerSISC2014} considered point-wise convergence of the mean and variance of a Gibbs SSOR sampler accelerated by Chebyshev polynomials.  In this paper we prove convergence in distribution for Gibbs samplers corresponding to {\it any} matrix splitting and accelerated  by {\it any} polynomial that is independent of the Gibbs iterations.  We then apply a polynomial accelerated sampler to solve a massive Bayesian linear inverse problem that is infeasible to solve using conventional techniques.

Chebyshev acceleration requires estimates of the extreme eigenvalues of the error operator, which we obtain via a conjugate-gradient (CG) algorithm at no significant computational cost \citep{Meurant}. The CG algorithm itself may be
adapted to sample from normal distributions; the CG solver and corresponding sampler,  depicted in panels E and F of
Figure~\ref{fig:quadraticANDGaussians}, were analysed by~\cite{ParkerFoxSISC2012} and is discussed in the supplementary material.

\zapthis{Chebyshev acceleration requires estimates of
the extreme eigenvalues of the operator that acts on error, which we obtain
via a conjugate-gradient (CG) algorithm at no significant increase in
computational cost. The CG algorithm itself may be
adapted to sample from normal distributions, giving an algorithm
that is direct (`perfect' in the language of statistics) in exact arithmetic \citep{FoxCD,ParkerFoxSISC2012}. The CG solver and corresponding sampler are depicted in panels E and F of
Figure~\ref{fig:quadraticANDGaussians}.
The direct/perfect nature of the algorithms is shown by convergence of the solver requiring a finite number of steps, and the sampler aligning with directions that are independent under the target distribution. However, that algorithm does not correspond to acceleration by a fixed polynomial,
and is outside the scope of this paper. The Chebyshev accelerated
SSOR sampler may also be further accelerated (significantly) by initializing with a CG sample, at no increase in computational cost. That idea is briefly outlined in section \ref{sect:acc_init}; an analysis of
the CG sampler was presented by \cite{ParkerFoxSISC2012}.}

\subsection{Structure of the paper}

In Section~\ref{sec:SampNorm} we review efficient methods for sampling from normal distributions,
highlighting Gibbs sampling in various algorithmic forms.
Standard results for stationary iterative solvers are presented in Section~\ref{sec:StatProc}.
Theorems in 
Section~\ref{sect:equiv} establish equivalence of convergence and
convergence factors for iterative solvers and Gibbs samplers.
Application of polynomial acceleration methods to linear solvers and Gibbs sampling is given
in Section~\ref{sec:NonStat}, including a proof of convergence of the first and second moments of a polynomial
accelerated sampler. Numerical verification
of convergence results is presented in
Section~\ref{sect:num}.

\section{Sampling from multivariate normal distributions}
\label{sec:SampNorm}
We consider the problem of sampling from an $n$-dimensional normal distribution $\Norm(\bmu,\bSigma)$ defined by the
mean $n$-vector $\bmu$, and the $n\times n$ symmetric and positive definite (SPD) covariance matrix $\bSigma$.
Since if $\mathbf{z}\sim \Norm(\mathbf{0},\bSigma)$ then $\mathbf{z}+\bmu\sim \Norm(\bmu,\bSigma)$, it often suffices to consider drawing samples from normal distributions with zero mean. An exception is when $\bmu$ is defined implicitly, which we discuss in section~\ref{sec:GeneralNoise}.

In Bayesian formulations of inverse problems that use a GMRF as a prior distribution, typically the precision matrix $\A=\bSigma^{-1}$ is explicitly modeled and available \citep{RueHeld,Higdon}, perhaps as part of a hierarchical model \citep{Bannerjee}. Typically then the precision matrix (conditioned on hyperparameters) is large though sparse, if the neighborhoods specifying conditional independence are small. We are particularly interested in this case, and throughout the paper will focus 
on sampling from $\Norm(\f{0},\A^{-1})$ when $\A$ is sparse and large, or when some other property makes operating by $\mathbf{A}$ easy, i.e.,
one can evaluate $\mathbf{Ax}$ for any vector $\mathbf{x}$.

Standard sampling methods for moderately sized normal distributions utilize the Cholesky factorization~\citep{Rue2,RueHeld}
since it is fast, incurring approximately $(1/3)n^3$ floating point operations (flops) and is backwards stable
\citep{Watkins}. Samples can also be drawn using the more expensive eigen-decomposition \citep{RueHeld}, that costs approximately $(10/3)n^3$ flops, or more generally using mutually conjugate vectors~\citep{FoxCD,ParkerFoxSISC2012}. For stationary Gaussian random fields defined on the lattice, Fourier methods can lead to efficient sampling for large problems~\citep{Gneiting}.

Algorithm~\ref{alg:CholCov} shows the steps for sampling from $\Norm(\f{0},\bSigma)$ using Cholesky factorization, when the
covariance  matrix $\bSigma$ is available~\citep{Neal,MacKay,Higdon}.\\
\begin{algorithm}[H]
\SetKwInOut{Input}{input}\SetKwInOut{Output}{output}
  \Input{Covariance matrix $\bSigma$}
  \Output{$\mathbf{y}\sim \Norm(\f{0},\bSigma)$}
  \BlankLine
  Cholesky factor $\bSigma=\mathbf{CC}^T$\;
  sample $\mathbf{z}~\sim~\Norm(\f{0},\mathbf{I})$\;
  $\mathbf{y}=\mathbf{Cz}$\;
  \caption{Cholesky sampling using a covariance matrix
  $\bSigma$}\label{alg:CholCov}
\end{algorithm}

When the precision matrix $\A$ is available, a sample $\mathbf{y}\sim
\Norm(\f{0},\A^{-1})$ may be drawn using Algorithm~\ref{alg:CholPrec} given by~\cite{Rue2}~\cite[see also][]{RueHeld}.\\
\begin{algorithm}[H]
\SetKwInOut{Input}{input}\SetKwInOut{Output}{output}
  \Input{Precision matrix $\A$}
  \Output{$\mathbf{y}\sim \Norm(\f{0},\A^{-1})$}
  \BlankLine
  Cholesky factor $\A=\mathbf{BB}^T$\;
  sample $\mathbf{z}~\sim~\Norm(\f{0},\mathbf{I})$\;
  solve $\mathbf{B}^T\mathbf{y=z}$ by back substitution\;
  \caption{Cholesky sampling using a precision matrix
$\A$}\label{alg:CholPrec}
\end{algorithm}
The computational cost of Algorithm~\ref{alg:CholPrec} depends on the bandwidth of
$\A$, that also bounds the bandwidth of the Cholesky
factor $\f{B}$. For a bandwidth $b$, calculation of the Cholesky
factorization requires $\mathcal{O}(b^2n)$ flops, which provides savings over the full-bandwidth case when $b\ll n/2$~\citep{Golub,Rue2,Watkins}.
For GMRF's defined over 2-dimensional domains, the use of a
bandwidth reducing permutation often leads to substantial
computational savings~\citep{Rue2,Watkins}.
In 3-dimensions and above, typically no
permutation exists that can significantly reduce the bandwidth below $n^{2/3}$, hence the cost of sampling by Cholesky factoring is at least $\mathcal{O}(n^{7/3})$ flops.
Further, Cholesky factorizing requires that the precision matrix and the Cholesky factor be stored in computer memory, which can be prohibitive for large problems.
In Section \ref{sect:num} we give an example of sampling from a large GMRF for which Cholesky factorization is prohibitively expensive.



\subsection{Gibbs sampling from a normal distribution}

Iterative samplers, such as 
Gibbs,  are an attractive option when drawing samples from high
dimensional multivariate normal distributions due to their inexpensive cost per
iteration  and small computer memory requirements (only
vectors of size $n$ need be stored).
If the precision
matrix is sparse with $\mathcal{O}(n)$ non-zero elements, then, regardless of
the bandwidth, iterative methods cost only about $2n$ flops per iteration, which
is comparable with sparse Cholesky factorizations.
However, when the bandwidth is $\mathcal{O}(n)$, the cost of the Cholesky factorization is high at
$\mathcal{O}(n^3)$ flops, while iterative methods maintain their
inexpensive cost per iteration. Iterative methods are then preferable when requiring
significantly fewer than $\mathcal{O}(n^2)$ iterations for adequate
convergence. 
In the examples presented in section~\ref{sect:num} we find that $\mathcal{O}(n)$ iterations give convergence to
machine precision, so the iterative methods are preferable for large problems.

\subsubsection{Componentwise formulation}

One of the simplest iterative sampling methods is the component-sweep Gibbs
sampler
\citep{Geman,Gelman,Gilks,RueHeld}.  Let $\mathbf{y}=(y_1, y_2, ..., y_n)^T\in\Re^n$ denote a
 vector in terms of its components, and let
$\mathbf{A}$ be an $n\times n$ precision matrix with
elements $\{a_{ij}\}$. One sweep over all $n$ components can be written as in
Algorithm~\ref{alg:Gibbs}~\citep{Barone}, showing that the algorithm can be implemented using vector and
scalar operations only, and storage or inversion of the
precision matrix $\mathbf{A}$ is not required.

\begin{algorithm}[H]
\SetKwInOut{Input}{input}\SetKwInOut{Output}{output} 
  \Input{Precision matrix $\A$, initial state $\f{y}^{(0)}=\left(y_1^{(0)}, y^{(0)}_2,...,y^{(0)}_n\right)^T$, and maximum iteration  $k_{\text{max}}$}
  \Output{$\left\{\f{y}^{(0)},\f{y}^{(1)},\f{y}^{(2)},\ldots,\f{y}^{(k_{\text{max}})}\right\}$ where $\f{y}^{(k)}\overset{\mathcal{D}}{\to}\Norm(\f{0},\f{A}^{-1})$ as $k\to\infty$}
  \BlankLine
  \For{$k=1,2,\ldots,k_{\mathrm{max}}$}
   {\For{ $i=1,2,\ldots,n$}
     {Sample $z \sim \Norm(0,1)$\;
      $y_{i}^{\left(k\right)  }=\displaystyle\frac{z}{\sqrt{a_{ii}}}
       - \frac{1}{a_{ii}}\left(\sum_{j>i} a_{ij}y_{j}^{\left(  k-1\right)  }
       -\sum_{j<i}a_{ij}y_{j}^{\left(  k\right)} \right) $
     }
   }
%
%
  \caption{Component-sweep Gibbs sampling using a precision matrix
$\A$}
  \label{alg:Gibbs}
\end{algorithm}
The index $k$ may be
omitted (and with `$=$' interpreted as assignment) to give an algorithm
that can be evaluated \emph{in place}, requiring minimal storage.
\zapthis{Variants of the Gibbs sampler use different choices of the
sequence of updating coordinates, rather than the regular sweep
$i=1,2,\ldots,n$ used above.
Then, the
conditions $j>i$ and $j<i$ are generalized to ``those components
updated in this sweep," and ``those components yet to be updated,"
respectively.}



\subsubsection{Matrix formulation}

One iteration in Algorithm~\ref{alg:Gibbs} consists of a sweep
over all $n$ components of $\mathbf{y}^{(k)}$ in sequence. The
iteration can be written succinctly in the matrix form \citep{GoodmanSokalMGMC}
    \be
    \label{Gibbs}
    \mathbf{y}^{(k+1)} = -\mathbf{D}^{-1}\mathbf{Ly}^{(k+1)}
                       - \mathbf{D}^{-1}\mathbf{L}^T\mathbf{y}^{(k)}
                       + \mathbf{D}^{-1/2}\mathbf{z}^{(k)},
    \ee
where $\mathbf{z}^{(k)}\sim \Norm(\f{0},\mathbf{I})$,
$\mathbf{D}={\rm diag}(\mathbf{A})$, and $\mathbf{L}$ is the
strictly lower triangular part of $\mathbf{A}$.  This equation
makes clear that the computational cost of each sweep is about
$2n^2$ flops, when $\f{A}$ is dense, due to multiplication by the triangular matrices
$\mathbf{L}$ and $\mathbf{L}^T$, and $\mathcal{O}(n)$ flops when $\f{A}$ is sparse.

Extending this formulation to sweeps over any other \emph{fixed}
sequence of coordinates is achieved by putting $\mathbf{P}\mathbf{A}\mathbf{P}^{\text{T}}$
in place of $\mathbf{A}$ for some permutation matrix
$\mathbf{P}$. The use of
\emph{random} sweep Gibbs sampling has also been suggested~\citep{AmitGrenander,Fishman,LiuWongKong95,RobertsSahu}, though we do not consider that here.

\subsubsection{Convergence}
If the iterates
$\mathbf{y}^{\left(  k\right)  }$ in~\eqref{Gibbs} converge in distribution to a distribution $\Pi$
which is independent of the starting state $\mathbf{y}^{\left(  0\right)  }$,
then the sampler is
\emph{convergent}, and we write
    $$\mathbf{y}^{\left(  k\right)  }\overset{\mathcal{D}}{\to}\Pi.$$
It is well known that the iterates  $\mathbf{y}^{(k)}$ in the Gibbs sampler~\eqref{Gibbs} converge in distribution geometrically
to $\Norm(\f{0},\mathbf{A}^{-1})=\Norm(\f{0},\bSigma)$  \citep{RobertsSahu}.  We
consider geometric convergence in detail in Section~\ref{sect:equiv}.

\section{Linear stationary iterative methods as linear equation solvers}
\label{sec:StatProc}
Our work draws heavily on existing results for stationary
linear iterative methods for solving linear systems. Here we briefly
review the main results that we will use.

Consider a system of linear equations written as the matrix
equation
\begin{equation}
  \mathbf{Ax}=\mathbf{b}
  \label{eqn:linear}
\end{equation}
where $\mathbf{A}$ is a given $n\times n$ nonsingular matrix and
$\mathbf{b}$ is a given $n$-dimensional vector. The problem is to
find an $n$-dimensional vector $\mathbf{x}$ satisfying equation
\eqref{eqn:linear}. Later we will consider the case where
$\mathbf{A}$ is symmetric positive definite (SPD) as holds for
covariance and precision matrices~\citep{Feller}.

\subsection{Matrix splitting form of stationary iterative algorithms}

A common class of methods for solving \eqref{eqn:linear} are the
linear iterative methods based on a \emph{splitting} of
$\mathbf{A}$ into $\mathbf{A}=\mathbf{M}-\mathbf{N}$. The matrix splitting is
the standard way of expressing and
analyzing linear iterative algorithms, following its introduction
by \cite{Varga}.
The system \eqref{eqn:linear} is then transformed to %
$
  \mathbf{M}\mathbf{x}=\mathbf{N}\mathbf{x}+\mathbf{b}%
$
or, if $\mathbf{M}$ is nonsingular, %
$
  \mathbf{x}=\mathbf{M}^{-1}\mathbf{Nx}+\mathbf{M}^{-1}\mathbf{b}.
$
The iterative methods use this equation to compute successively better
approximations $\mathbf{x}^{\left(  k\right)  }$ to the solution using the
iteration step
\begin{align}
  \x^{\left(  k+1\right)  }  &  =\M^{-1}\N\x^{\left(
  k\right)  }+\mathbf{M}^{-1}\mathbf{b}\label{eqn:iteration}
    =\mathbf{Gx}^{\left(  k\right)  }+\mathbf{g}. 
\end{align}

We follow the standard terminology used for these methods
\cite[see e.g.][]{Ax,Golub,SaadIter,Young}.
Such methods are
termed linear stationary iterative methods (of the first degree);
they are stationary\footnote{This use of \emph{stationary} corresponds to the term \emph{homogeneous} when referring to a Markov chain. It is not to be confused with a \emph{stationary distribution} that is invariant under the iteration. Later we will develop non-stationary iterations, inducing a non-homogeneous Markov chain that will, however, preserve the target distribution at each iterate.} because the \emph{iteration matrix}
$\mathbf{G}=\mathbf{M}^{-1}\mathbf{N}$ and the vector
$\f{g}=\mathbf{M}^{-1}\mathbf{b}$ do not depend on $k$.  The splitting
is \emph{symmetric} when both $\mathbf{M}$ and $\mathbf{N}$ are
symmetric matrices. The iteration, and splitting, is
\emph{convergent} if $\mathbf{x}^{\left(  k\right) }$ tends to a
limit independent of $\mathbf{x}^{(0)}$, the limit being $\A^{-1}\f{b}$ (see,
e.g.~\cite[Theorem
5.2]{Young}).

The iteration~\eqref{eqn:iteration} is often written in the residual
form
so that convergence may be monitored in terms of the norm of the
residual vector, and emphasizes that $\mathbf{M}^{-1}$ is acting
as an approximation to $\mathbf{A}^{-1}$, as in
Algorithm~\ref{alg:iter}.\\
\begin{algorithm}[H]
\SetKwInOut{Input}{input}\SetKwInOut{Output}{output} 
  \Input{The splitting $\M$, $\N$ of $\A$, initial state $\f{x}^{(0)}$}
  \Output{$\x^{(k)}$ approximating $\x^{\ast}=\A^{-1}\mathbf{b}$}
  \BlankLine
   $k=0$\;
   \Repeat
     { $\left\| \f{r}^{\left(k\right)  }\right\| $ is sufficiently small}
     {$\mathbf{r}^{\left( k\right)}=\mathbf{b}-\mathbf{Ax}^{\left(k\right)  }$\;
      $\x^{\left(  k+1\right)  }=\x^{\left(  k\right)  }+\M^{-1}\f{r}^{\left(
k\right)}$\;
      increment $k$\;
      \BlankLine
      }
%
%
  \caption{Stationary iterative solve of $\mathbf{Ax=b}$}
  \label{alg:iter}
\end{algorithm}


In computational algorithms it is important to note that
the symbol $\mathbf{M}^{-1}\mathbf{r}$ is interpreted as ``solve the system
$\mathbf{Mu}=\mathbf{r}$ for $\mathbf{u}$" rather than ``form the inverse of
$\mathbf{M}$ and multiply $\mathbf{r}$ by $\mathbf{M}^{-1}$" since the latter is
much more
computationally expensive (about $2n^3$ flops \citep{Watkins}). Thus, the practicality
of a splitting depends on the ease with which one can solve
$\mathbf{Mu}=\mathbf{r}$ 
for any vector $\mathbf{r}$.

\subsubsection{The Gauss-Seidel algorithm}
Many splittings of the matrix $\mathbf{A}$ use the terms in the expansion
$\mathbf{A}=\mathbf{L}+\mathbf{D}+\mathbf{U}$ where $\mathbf{L}$ is the
strictly lower triangular part of $\mathbf{A}$, $\mathbf{D}$ is the diagonal
of $\mathbf{A}$, and $\mathbf{U}$ is the strictly upper triangular part.

For
example, choosing $\mathbf{M}=\mathbf{L}+\mathbf{D}$ (so $\mathbf{N}%
=-\mathbf{U}$) allows 
$\mathbf{Mu}=\mathbf{r}$
to be solved by ``forward
substitution" (at a cost of $n^2$ flops when $\f{A}$ is dense), and hence does not require inversion or
Gauss-elimination of $\mathbf{M}$ (which would cost $2/3n^3$ flops when $\f{A}$ is dense). Using
this splitting in Algorithm~\ref{alg:iter}  results in the
\emph{Gauss-Seidel} iterative algorithm. When $\mathbf{A}$ is symmetric,
$\mathbf{U=L}^{\text{T}}$, and the Gauss-Seidel iteration can be
written as
\be \label{GS}
  \mathbf{x}^{\left(  k+1\right) }=-\mathbf{D}^{-1}\mathbf{Lx}^{\left(
  k+1\right) }-\mathbf{D}^{-1}\mathbf{L}^{\text{T}}\mathbf{x}^{\left(
  k\right)}+\mathbf{D}^{-1}\mathbf{b}.
\ee%
Just as we pointed out for the Gibbs sampler, variants of the
Gauss-Seidel algorithm such as ``red-black"
coordinate updates \citep{SaadIter}, may be written in this form using a suitable
permutation matrix.

The component-wise form of the Gauss-Seidel
algorithm can be written in `equation' form just as the Gibbs sampler  \eqref{Gibbs} was in Algorithm \ref{alg:Gibbs}.
The component-wise form emphasizes that Gauss-Seidel can be
implemented using vector and scalar operations only, and neither
storage nor inversion of the splitting is required.



\subsection{Convergence}
\label{sect:conv} A fundamental theorem of linear stationary
iterative methods states that the splitting
$\mathbf{A}=\mathbf{M}-\mathbf{N}$, where $\M$ is nonsingular, is
convergent (i.e., $\x^{(k)}\to \A^{-1}\mathbf{b}$ for any $\x^{(0)}$)
if and only if $\varrho\left( \M^{-1}\mathbf{N}\right) <1$, where
$\varrho\left( \cdot\right)  $ denotes the spectral radius of a
matrix~\cite[Theorem 3.5.1]{Young}. This characterization is often
used as a definition \citep{Ax,Golub,SaadIter}.

The \emph{error} at step $k$ is $\e^{(k+1)}=\x^{(k+1)} -\x^{\ast}$,
where $\x^{\ast}=\A^{-1}\mathbf{b}$. It follows that
\begin{equation}
  \e^{(k+1)}=(\M^{-1}\N)^k\e^{(0)}
  \label{eq:error}
\end{equation}
and hence the asymptotic average reduction in error per iteration is the multiplicative factor
\begin{equation}
\label{eqn:varrho}
\lim_{k\to\infty}\left(\frac{||\e^{(k+1)}||_2}{||\e^{(0)}||_2}\right)^{1/k}
=\varrho(\f{M}^{-1}\f{N})
\end{equation}
\cite[p. 166]{Ax}.  In numerical
analysis this is called the (asymptotic average) \emph{convergence factor}
\citep{Ax,SaadIter}. Later, we will show that this is exactly
the same as the quantity called the geometric \emph{convergence rate} in the
statistics literature \cite[see e.g.][]{RobertCasella}, for the equivalent Gibbs
sampler. We will use the term `convergence factor' throughout this paper to
avoid a clash of terminology, since in numerical analysis the \emph{rate} of
convergence is minus the log of the convergence factor \cite[see e.g.][p.
166]{Ax}.

%

\subsection{Common matrix splittings}
We now present the matrix splittings corresponding to  some common stationary linear iterative
solvers, with details for the case where $\A$ is symmetric, as holds for precision or
covariance matrices.

We have seen that the Gauss-Seidel iteration uses the splitting
$\M_{\text{GS}}=\f{L}+\D$ and $N_{\text{GS}}=-\f{L}^T$.
Gauss-Seidel is one of the simplest splittings and solvers, but is
also quite slow. Other splittings 
have
been developed, though the speed of each method is often problem
specific. Some common splittings are shown in
Table~\ref{tab:splittings}, listed with, roughly, greater speed
downwards. Speed of convergence in a numerical example is presented later in
Section~\ref{sect:num}.
\begin{table}
\caption{\label{tab:splittings}Common stationary linear solvers generated by splittings $\A=\M-\N$,
and conditions that guarantee convergence when $\A$ is SPD}
\centering
\fbox{%
\begin{tabular}[c]{|r|c|c|}
splitting & $\mathbf{M}$ & convergence \\
\hline
Richardson (R) & $\frac{1}{\omega}\mathbf{I}$ & $\displaystyle 0<\omega<\frac{2}{\varrho(\mathbf{A})}$ \\
Jacobi (J) & $\mathbf{D}$ & $\f{A}$ strictly diagonally dominant\\
Gauss-Seidel (GS) & $\mathbf{D+L}$ & always \\
SOR & $\frac{1}{\omega}\mathbf{D}+\mathbf{L}$ & $0<\omega<2$ \\
SSOR & $\frac{\omega}{2-\omega}\mathbf{M}_{\text{SOR}}\mathbf{D}^{-1}\mathbf{M}_{\text{SOR}}^{\text{T}}$ & $0<\omega<2$ \\
\end{tabular}}
\end{table}




The method of {\em successive over-relaxation} (SOR) uses the
splitting
    \be
    \label{SORsplit}
    \M _{\text{SOR}}=\frac{1}{\omega}\mathbf{D+L} ~~{\rm and}~~
\mathbf{N}_{\text{SOR}}=\frac{1-\omega}{\omega}\mathbf{D-L}^T
    \ee
in which $\omega$ is
 a \emph{relaxation} parameter chosen with $0<\omega< 2$.   SOR with $\omega=1$
is Gauss-Seidel. For optimal
values of $\omega$ such that
$\varrho(\M^{-1}_{\text{SOR}}\mathbf{N}_{\text{SOR}})<\varrho(\M^{-1}_{\text{GS}}
\mathbf{N}_{\text{GS}})$, SOR is an accelerated Gauss-Seidel iteration.
Unfortunately, there is no closed form for the optimal value of
$\omega$ for an arbitrary matrix $\mathbf{A}$, and the interval of
values of $\omega$ which admits accelerated convergence can be quite narrow
\citep{Young,Golub,SaadIter}.

The {\em symmetric-SOR} method (SSOR) incorporates both a forward and
backward sweep of SOR so that if $\mathbf{A}$ is symmetric then
the splitting
is symmetric \citep{Golub,SaadIter},
    \be
    \label{SSORsplit}
    \M _{\text{SSOR}}=\frac{\omega}{2-\omega}\M _{\text{SOR}}\mathbf{D}^{-1}\M _{\text{SOR}}^T
~~{\rm and}~~
\mathbf{N}_{\text{SSOR}}=\frac{\omega}{2-\omega}\mathbf{N}_{\text{SOR}}^T\mathbf{D}^{-1}
\mathbf{N}_{\text{SOR}}.
\ee
We will make use of symmetric splittings in conjunction with
polynomial acceleration in Section~\ref{sec:NonStat}.

When the matrix $\A$ is dense, 
Gauss-Seidel and SOR cost about $3n^2$ flops per iteration, with
$2n^2$ due to multiplication by the matrix $\mathbf{A}$ (in
order to calculate the residual) and another $n^2$ for the
forward substitution to solve $\mathbf{Mu=r}$.  Richardson incurs
no cost to solve $\mathbf{Mu=r}$, while a solve with the diagonal
Jacobi matrix incurs $n$ flops. Iterative methods are particularly attractive
when the matrix $\A$ is sparse, since then the cost per iteration is only
$O(n)$ flops.

Many theorems establish convergence of splittings by utilizing
properties of $\mathbf{A}$ in specific applications. Some general
conditions that guarantee convergence when $\mathbf{A}$ is SPD are
given in the right column of Table~\ref{tab:splittings}
\citep{Golub,SaadIter,Young}.

%

\section{Equivalence of stationary linear solvers and Gibbs samplers}
\label{sect:equiv}
We first consider the equivalence between linear solvers and stochastic iterations
in the case where the starting state and noise are not
necessarily normally distributed, then in Section~\ref{sec:SampNormSplitting}
\emph{et seq.} we restrict consideration to normal distributions.

\subsection{General noise}
\label{sec:GeneralNoise}
The striking similarity between the Gibbs sampler~\eqref{Gibbs} and the Gauss-Seidel iteration~\eqref{GS} is no coincidence.  It is an example of a general equivalence between the stationary linear solver derived from a splitting and the associated stochastic iteration used as a sampler.  We will make explicit the equivalence in the following theorems and corollary.
In the first theorem we show that a splitting is convergent (in the sense of stationary iterative solvers) if and
only if the associated stochastic iteration is convergent in distribution.


\begin{theorem}
\label{thm:StatConvSampConv} Let $\mathbf{A=M-N}$ be a splitting with $\M$ invertible, and let $\pi\left(\mathbf{\cdotp}\right)$ be some fixed probability distribution with zero mean and fixed non-zero covariance. For any fixed vector $\mathbf{b}$, and random vectors $\mathbf{c}^{\left(k\right)}\overset{\text{iid}}{\sim}\pi$, $k=0,1,2,\ldots$, the stationary linear
iteration
\begin{equation}
  \mathbf{x}^{\left(  k+1\right)  }=\M^{-1}\mathbf{Nx}^{\left(
  k\right)  }+\M^{-1}\mathbf{b}
  \label{eqn:solver}
\end{equation}
converges, with $\mathbf{x}^{\left(  k\right)  }\rightarrow\mathbf{A}%
^{-1}\mathbf{b}$ as $k\rightarrow\infty$ whatever the initial vector
$\x^{(0)}$, if and only if there exists a distribution $\Pi$ such that the stochastic iteration%
\begin{equation}
 \mathbf{y}^{\left(  k+1\right)  }=\M^{-1}\mathbf{Ny}^{\left(
  k\right)  }+\M^{-1}\mathbf{c}^{\left(  k\right)  }
\label{eq:sampler}
\end{equation}
converges in distribution to $\Pi$, with $\mathbf{y}^{\left(  k\right)
}\overset{\mathcal{D}}{\rightarrow}\Pi$ as $k\rightarrow\infty$ whatever the
initial state $\y^{(0)}$.
\end{theorem}
\proof  If the linear iteration~\eqref{eqn:solver} converges, then
$\varrho(\f{M}^{-1}\f{N})<1$ \cite[Thm 3-5.1 in][]{Young}. Hence there exists a unique distribution $\Pi$ with
$\mathbf{y}^{\left(k+1\right)}\overset{\mathcal{D}}{\rightarrow}\Pi$ \cite[Theorem 2.3.18-4 of][]{Duflo}, which shows necessity.
Conversely, if the linear solver does not converge to a limit independent of $\x^{(0)}$ for some $\mathbf{b}$, that also holds for $\mathbf{b}=\f{0}$ and hence initializing the sampler with $\E\left[\mathbf{y}^{(0)}\right]=\mathbf{x}^{(0)}$ yields
$\E\left[\mathbf{y}^{(k+1)}\right] = (\M^{-1}\mathbf{N})^k\x^{(0)}$ which does not converge to a value independent of $\f{y}^{(0)}$. Sufficiency holds by the contrapositive.  \hfill $\Box$

Convergence of the stochastic iteration~\eqref{eq:sampler} could also be established via the more general theory of \cite{DiaconisFreedman} that allows the iteration operator $\f{G}=\f{M}^{-1}\f{N}$ to be random, with convergence in distribution guaranteed when $\f{G}$ is \emph{contracting on average}; see \cite{DiaconisFreedman} for details.


The following theorem shows how to design the noise distribution $\pi$ so that the limit distribution $\Pi$ has a desired mean $\mu$ and covariance $\bSigma=\A^{-1}$.

\begin{theorem}
\label{thm:momentconv} Let $\A$ be SPD, $\mathbf{A=M-N}$ be a convergent
splitting, $\bmu$ a fixed vector, and $\pi\left(\mathbf{\cdotp}\right)$ a
fixed probability distribution with finite mean $\nu$ and non-zero covariance $\f{V}$.
Consider the stochastic iteration~\eqref{eq:sampler} where
$\f{c}^{(k)}\overset{iid}{\sim}\pi$, $k=0,1,2,\ldots$. Then, whatever the
starting state $\mathbf{y}^{(0)}$, the following are equivalent:
\begin{enumerate}
  \item $\E\left[\f{c}^{(k)}\right]=\bnu$
    and $\Var\left(\f{c}^{(k)}\right)=\f{V}=\f{M}^T + \f{N}$
  \item the iterates $\f{y}^{(k)}$ converge in distribution to some
distribution $\Pi$ that has mean $\bmu=\f{A}^{-1}\nu$ and covariance matrix
$\f{A}^{-1}$; in particular $\E\left[\f{y}^{(k)}\right]\to\bmu$
    and $\Var\left(\f{y}^{(k)}\right)\to\f{\A}^{-1}$ as
$k\to\infty$.
\end{enumerate}
\end{theorem}

\proof Appendix~\ref{sect:ap:VarB}. $\Box$

Additionally, the mean and covariance converge geometrically, with convergence factors given by the
convergence factors for the linear iterative method, as established in the following corollary.
\begin{corollary}
\label{cor:StatConvSampConv}
The first and second moments of iterates in the stochastic iteration in Theorem~\ref{thm:momentconv} converge geometrically.  Specifically,
$\E\left(\mathbf{y}^{(k)}\right) \to \bmu$ with convergence factor
$\varrho(\M^{-1}\mathbf{N})$ and $\Var\left(\mathbf{y}^{(k)}\right)=\f{A}^{-1} -
\f{G}^k\left(\f{A}^{-1}-\Var(\f{y}^{(0)})\right)(\f{G}^k)^T
\to\mathbf{A}^{-1}$ with convergence factor $\varrho(\M^{-1}\mathbf{N})^2$.
\end{corollary}
\proof Appendix~\ref{sect:ap:VarB}. $\Box$

Note that the matrix splitting has allowed an explicit construction of the noise covariance to give a desired precision matrix of the target distribution. We see from Theorem~\ref{thm:momentconv} that the stochastic iteration may be designed to converge to a distribution with  non-zero target mean, essentially by adding the deterministic iteration~\eqref{eqn:solver} to the stochastic iteration~\eqref{eq:sampler}. This is particularly useful when the mean is defined implicitly via solving a matrix equation. In cases where the mean is known explicitly, the mean may be added after convergence of the stochastic iteration with zero mean, giving an algorithm with faster convergence since the covariance matrix converges with factor $\varrho(\M^{-1}\mathbf{N})^2<\varrho(\M^{-1}\mathbf{N})$ \cite[this was also noted by][]{Barone2}. Convergence in variance for non-normal targets was considered in~\cite{FoxParkerSISC2014}.

Using Theorems~\ref{thm:StatConvSampConv}  and \ref{thm:momentconv}, and Corollary~\ref{cor:StatConvSampConv} we can draw on the vast literature in numerical linear algebra on
stationary linear iterative methods to find random
iterations that are computationally efficient and provably
convergent in distribution with desired mean and covariance.
In particular, results in \cite{AmitGrenander,Barone,GalliGao}, \cite{RobertsSahu}, and \cite{LiuWongKong95} are special cases of the general theory of matrix splittings presented here.

\subsection{Sampling from normal distributions using matrix splittings}
\label{sec:SampNormSplitting}
We now restrict attention to the case of normal target distributions.


\begin{corollary}
\label{cor:normsampler} If in Theorem~\ref{thm:momentconv} we set
$\pi=\Norm(\bnu,\f{V})$, for some non-zero covariance matrix $\f{V}$, then, whatever the starting state $\f{y}^{(0)}$, the
following are equivalent:  (i) $\f{V}=\f{M}^T+\f{N}$; (ii) $\f{y}^{(k)}\overset{\mathcal{D}}{\to}\Norm(\bmu,\mathbf{A}^{-1})$ where $\bmu=\f{A}^{-1}\nu$.
\end{corollary}
\proof Since $\pi$ is normal, then $\Pi$ in Theorem~\ref{thm:momentconv} is normal.
Since a normal
distribution is sufficiently described by its first two moments, the corollary
follows.
\hfill $\Box$


Using Corollary~\ref{cor:normsampler}, we found normal sampling
algorithms corresponding to some common stationary linear solvers.
The results  are given in Table~\ref{tab:statsamp}.
\begin{table}
\caption{\label{tab:statsamp}Some generalized Gibbs samplers for drawing from $\Norm\left(\f{0},\f{A}^{-1}\right)$ adapted from common stationary linear solvers. Each Gibbs iteration requires sampling the noise vector $\f{c}^{(k)}\sim
\Norm\left(\f{0},\f{M}^T+\f{N}\right)$
}

\centering
\fbox{%
\begin{tabular}{|r|c|c|}
  \hline
  Sampler & $\M $ & $\Var\left(\f{c}^{(k)}\right)=\f{M}^T+\f{N}$\\
  \hline
   Richardson & $\displaystyle\frac{1}{\omega}\mathbf{I}$ &
$\displaystyle\frac{2}{\omega}\mathbf{I}-\mathbf{A}$\\[0.5em]
   Jacobi & $\mathbf{D}$ & $2\f{D}-\mathbf{A}$\\[0.5em]
 Gibbs (Gauss-Seidel) & $\mathbf{D+L}$ & $\f{D}$\\[0.5em]
  SOR & $\displaystyle\frac{1}{\omega} \f{D + L}$ & $\displaystyle\frac{2-\omega}{\omega} \f{D}$\\[0.5em]
  SSOR (REGS) & $\displaystyle\frac{\omega}{2-\omega}\M
_{\text{SOR}}\f{D}^{-1}\M _{\text{SOR}}^T$ & $\displaystyle\frac{\omega}{2-\omega}\left(\M
_{\text{SOR}}\f{D}^{-1}\M _{\text{SOR}}^T +
\mathbf{N}_{\text{SOR}}^T\f{D}^{-1}\mathbf{N}_{\text{SOR}}\right)$\\[0.5em]
  \end{tabular}}
\end{table}
A sampler corresponding to a convergent splitting is implemented in Algorithm~\ref{alg:sampler}.\\
\begin{algorithm}[H]
\SetKwInOut{Input}{input}\SetKwInOut{Output}{output}
  \Input{SPD precision matrix $\A$, $\M$ and $\N$ defining a convergent
splitting of $\A$, number of steps $k_{\max}$, initial
state $\y^{(0)}$}
  \Output{$\mathbf{y}^{(k)}$ approximately distributed as $\Norm(\f{0},\A^{-1})$}
  \BlankLine
   \For
     { $k=0,\ldots,k_{\max}$}
     {sample $\f{c}^{(k)}\sim \Norm(\f{0},\f{M}^{T}+\f{N})$\;
      $\f{y}^{(k+1)} = \M^{-1}(\f{N}\y^{(k)} + \f{c}^{(k)})$
      }
  \caption{Stationary sampler of
$\Norm(\f{0},\mathbf{A}^{-1})$}\label{alg:sampler}
\end{algorithm}

\noindent The assignment $\f{y}^{(k+1)} = \M^{-1}(\f{N}\y^{(k)} + \f{c}^{(k)})$ in
Algorithm~\ref{alg:sampler}  can be replaced by the slightly more expensive
steps  $\f{r}^{(k)}=\f{c}^{(k)}-\A\y^{(k)}$ and
$\mathbf{y}^{(k+1)}=\mathbf{y}^{(k)} + \M^{-1}\f{r}^{(k)}$, which allows
monitoring
of the residual, and emphasizes the equivalence with the
stationary linear solver in Algorithm~\ref{alg:iter}.
Even though
convergence may not be diagnosed by a decreasing norm of the residual, lack
of convergence can be diagnosed when the residual diverges in magnitude. In practice, 
the effective convergence factor for a sampler may be calculated by solving the linear system~\eqref{eqn:linear} (perhaps with a random right hand side) using the iterative solver derived from the splitting and monitoring the decrease in error to evaluate the asymptotic average convergence factor using equation \eqref{eqn:varrho}. By Corollary~\ref{cor:StatConvSampConv}, this estimates the convergence factor for the sampler.

The practicality of a sampler derived from a convergent splitting depends on the ease with which one can solve $\M\y=\f{r}$ for any $\f{r}$ (as for the stationary linear solver) and also the ease of drawing iid noise vectors from $ \Norm(\f{0},\f{M}^{T}+\f{N})$.
Sampling the noise vector is simple when a matrix square root, such as the Cholesky factorization, of $\f{M}^{T}+\f{N}$ is cheaply available. Thus, a sampler is at least as expensive as the corresponding linear solver since, in addition to operations in each iteration, the sampler must factor the $n\times n$ matrix $\f{Var(c}^{(k)})=\f{M}^T+\f{N}$.
For the samplers listed in Table~\ref{tab:statsamp} it is interesting that the simpler the splitting, the more complicated is the variance of the noise. Neither Richardson nor Jacobi splittings give useful sampling algorithms since the difficulty of sampling the noise vector is no less than the original task of sampling from $ \Norm(\f{0},\f{A}^{-1})$. The Gauss-Seidel splitting, giving the usual Gibbs sampler, is at a kind of sweet spot, where solving equations in $\M$ is simple while the required noise variance is diagonal, so posing a simple sampling problem.

The SOR stationary sampler uses the SOR splitting $\f{M}_{\text{SOR}}$ and $\f{N}_{\text{SOR}}$ in \eqref{SORsplit}
for $0< \omega< 2$ and the noise vector $\f{c}^{(k)}\sim \Norm(\f{0},\f{M}_{SOR}^T+\f{N}_{SOR}=\frac{2-\omega}{\omega} \f{D})$ (Table~\ref{tab:statsamp}).
This sampler was introduced by~\cite{Adler}, rediscovered by~\cite{Barone}, and has been studied extensively
\citep{%
Barone2,GalliGao,LiuWongKong95,Neal2,RobertsSahu}.
For
$\omega=1$, the SOR sampler is a Gibbs (Gauss-Seidel) sampler.
For values of $\omega$ such that
$\varrho(\M_{\text{SOR}}^{-1}\N_{\text{SOR}})<\varrho(\M^{-1}_{GS}\N_{GS})$), the SOR-sampler is an
accelerated Gibbs sampler.  As for the linear solver, implementation of the Gibbs and SOR
samplers by Algorithm~\ref{alg:sampler} requires multiplication by
the upper triangular $\f{N}$ and forward substitution with respect
to $\f{M}$ at a cost of $2n^2$ flops.   In addition, these samplers must take the square root
of the diagonal matrix $\frac{2-\omega}{\omega}
\f{D}$ at a mere cost of $O(n)$ flops.

Implementation of an SSOR sampler instead of a Gibbs or SOR sampler is advantageous
since the Markov chain $\{\f{y}^{(k)}\}$ is reversible
\citep{RobertsSahu}.  SSOR sampling uses the symmetric-SOR splitting $\f{M}_{\text{SSOR}}$ and $\f{N}_{\text{SSOR}}$ in \eqref{SSORsplit}.
The SSOR stationary sampler is most easily implemented by forward and backward
SOR sampling sweeps as in Algorithm~\ref{alg:SSOR}, so the matrices
$\f{M}_{\text{SSOR}}$ and $\f{N}_{\text{SSOR}}$ need never be explicitly formed.\\
\begin{algorithm}[H]
\label{alg:SSORsampler}
\SetKwInOut{Input}{input}\SetKwInOut{Output}{output}
  \Input{The SOR splitting $\f{M},~ \f{N}$ of $\f{A}$, relaxation parameter $\omega$, initial state $\f{y}$, $k_{\max}$}
  \Output{$\y$ approximately distributed as $\Norm(\f{0,A}^{-1})$}
  \BlankLine
  set $\gamma=\left(\frac{2}{\omega}-1\right)^{1/2}$\;
  \BlankLine
   \For
     { $k=1,\ldots,k_{\max}$}
     {sample $\f{z} \sim \Norm(\f{0},\f{I})$\;
      $\f{x}:=\f{M}^{-1}(\f{Ny} + \gamma \f{D}^{1/2}\f{z})$\;
      sample $\f{z} \sim \Norm(\f{0},\f{I})$\;
      $\f{y} := \f{M}^{-T}(\f{N}^T\f{x} + \gamma \f{D}^{1/2}\f{z})$
      }
  \caption{SSOR sampling from $\Norm(\f{0,A}^{-1})$}
  \label{alg:SSOR}
\end{algorithm}

We first encountered restricted versions of Corollary \ref{cor:normsampler}
for normal
distributions in \cite{AmitGrenander} and in \cite{Barone} where geometric
convergence of the covariance matrices was established for the
Gauss-Seidel and SOR splittings.  These and the SSOR splitting
 were investigated in \cite{RobertsSahu} (who labelled the sampler REGS).

Corollary~\ref{cor:normsampler} and Table~\ref{tab:splittings}
show that the Gibbs,
SOR and SSOR samplers converge for any SPD precision matrix
$\mathbf{A}$.  This summarizes results in
\cite{Barone,GalliGao} and the deterministic sweeps investigated
in \cite{AmitGrenander,RobertsSahu,LiuWongKong95}.
Corollary~\ref{cor:normsampler} generalizes these
results 
for {\em
any} matrix splitting $\f{A=M-N}$ by guaranteeing convergence of the
random iterates \eqref{eq:sampler} to $\Norm(\f{0},\f{A}^{-1})$ with convergence factor
$\varrho(\f{M}^{-1}\f{N})$ (or $\varrho(\f{M}^{-1}\f{N})^2$ if $\mu=\f{0}$).

\section{Non-stationary iterative methods}
\label{sec:NonStat}
\subsection{Acceleration of linear solvers by polynomials}
\label{sec:poly}
A common scheme in numerical linear algebra for
accelerating a stationary method when $\M$ and $\mathbf{N}$ are
symmetric is through the use of polynomial preconditioners
\citep{Ax,Golub,SaadIter}.   Equation~\eqref{eq:error} shows
that after $k$ steps the error in the stationary method is a $k^{th}$ order
polynomial of the matrix $\f{I}-\mathbf{G}=\M^{-1}\f{A}$.
The idea behind polynomial acceleration is to implicitly implement a
different $k^{th}$ order polynomial $P_k(\M^{-1}\f{A})$ such that
$\varrho(P_k(\M^{-1}\f{A}))<\varrho(\left(\f{I}-\M^{-1}\f{A}\right)^k)$.  The coefficients of $P_k(\M^{-1}\f{A})$
are functions of a set of \emph{acceleration parameters}
$\{\{\alpha_k\},\{\tau_k\}\}$, introduced by the second order iteration
    \be
    \label{AccScheme}
    \mathbf{x}^{(k+1)}=(1-\alpha_k)\mathbf{x}^{(k-1)} + \alpha_k\mathbf{x}^{(k)} +
\alpha_k\tau_k \M^{-1}(\mathbf{b}-\mathbf{Ax}^{(k)}).
    \ee
At the first step, $\alpha_0=1$ and $\mathbf{x}^{(1)}=\mathbf{x}^{(0)} +
\tau_0 \M^{-1}(\mathbf{b}-\mathbf{Ax}^{(0)})$.
Setting $\alpha_k=\tau_k=1$ for every $k$ yields a basic un-accelerated
stationary method.  The accelerated iteration in \eqref{AccScheme} is implemented at a negligible increase in cost of $\mathcal{O}(n)$ flops per iteration (due to scalar-vector multiplication and vector addition) over the corresponding stationary solver \eqref{eqn:iteration}.

It can be shown that (e.g., \cite{Ax}) that the $(k+1)^{st}$ order polynomial $P_{k+1}$ generated recursively by the second order non-stationary linear solver \eqref{AccScheme} is
\be
\label{RecursionErrCheby}
P_{k+1}\left(\lambda\right)  =\left(  \alpha_{k}-\alpha_k\tau_k\lambda\right)  P_{k}\left(
\lambda\right)  +\left(  1-\alpha_{k}\right)  P_{k-1}\left(  \lambda\right).
\ee
This polynomial acts on the error $\f{e}^{(k)}=\f{x}^{(k)} - \A^{-1}\f{b}$ by
    $\f{e}^{(k+1)}=P_k(\M^{-1}\A)\f{e}^{(0)}$,
which can be compared directly to \eqref{eq:error}.

When estimates of the extreme eigenvalues $\lambda_{\min}$ and $\lambda_{\max}$ of
$\mathbf{I}-\mathbf{G}=\M^{-1}\mathbf{A}$ are available ($\lambda_{\min}$
and $\lambda_{\max}$ are real when $\M $ and $\mathbf{N}$ are symmetric), then the coefficients $\{\tau_k,\alpha_k\}$ can be chosen to generate the \emph{scaled Chebyshev} polynomials $\{Q_k\}$, which give optimal error reduction at every step.
The Chebyshev acceleration
parameters are
    \be
    \label{ChebyAccParams}
    \tau_k = \frac{2}{\lambda_{\max}+\lambda_{\min}}, ~~
        \beta_k=\left(\frac{1}{\tau_k}-\beta_{k-1}\left(\frac{\lambda_{\max}-\lambda_{\min}}{4}\right)^2\right)^{-1},~~
        \alpha_k=\frac{\beta_k}{\tau_k},
    \ee
where $\alpha_0=1$ and $\beta_0=\tau_0$ \citep{Ax}.  Note
that these parameters are independent of the
iterates $\{\mathbf{x}^{(k)}\}$.  Since $\M $ is required to be symmetric,
applying Chebyshev acceleration to SSOR is a common pairing; its effectiveness as a linear solver is shown later in Table~\ref{tab:opt}.

Whereas the stationary methods converge with asymptotic average convergence
factor $\varrho(\M^{-1}\N)$, the convergence factor for the Chebyshev method
depends on $\operatorname{cond}(\M^{-1}\A)=\lambda_{\max}/\lambda_{\min}$.
Specifically the scaled Chebyshev polynomial $Q_k(\f{\lambda})$ minimizes
$\max_{\lambda\in [ \lambda_{\min},\lambda_{\max} ]}P_k(\lambda)$ over all
$k^{th}$ order polynomials $P_k$, with
    \be
    \label{ChebyPolyOptimal}
    \max_{\lambda\in [ \lambda_{\min},\lambda_{\max} ]}|Q_k(\lambda)|=\frac{2\sigma^k}{1 + \sigma^{2k}}.
    \ee
Since the error at the $k^{th}$ step of a Chebyshev accelerated linear solver is $\f{e}^{(k+1)}=Q_{k}(\lambda)\f{e}^{(0)}$, then the asymptotic convergence factor is bounded above by
    \be
    \label{ChebyAccFactor}
    \sigma=\frac{1-\sqrt{\lambda_{\min}/\lambda_{\max}}}{1+\sqrt{\lambda_{\min}/\lambda_{\max}}}
    \ee
\cite[p.181][]{Ax}. Since $\sigma\in[0,1)$, the polynomial accelerated scheme is guaranteed to converge even if the original splitting was not convergent. Further, the convergence factor of the stationary iterative solver is bounded below by
\(
    \rho=\frac{1-{\lambda_{\min}/\lambda_{\max}}}{1+{\lambda_{\min}/\lambda_{\max}}}
    \)
\citep[see e.g.][Thm 5.9]{Ax}. Since $\sigma<\rho$ (except when $\lambda_{\min}=\lambda_{\max}$ in which case $\sigma=0$), polynomial acceleration always reduces the convergence factor, so justifies the term \emph{acceleration}.
The Chebyshev accelerated iteration~\eqref{AccScheme} is amenable to
preconditioning that reduces the condition number, and hence reduces $\sigma$, such as incomplete Cholesky
factorization or graphical methods~\citep{Ax,SaadIter}.  Axelsson also shows that after
\begin{equation}
\label{eq:secondordersolveriterations}
k^* = \lceil\frac{\ln (\varepsilon/2)}{\ln\sigma}\rceil
\end{equation}
iterations of the Chebyshev solver, the error reduction is
$||\e^{(k^*)}||_{\A^\nu}/||\e^{(0)}||_{\A^\nu}\le\varepsilon$
for some real number $\nu$ and any $0<\varepsilon < 1$ \cite[eqn 5.32]{Ax}.

\subsection{Acceleration of Gibbs sampling by polynomials}
\label{sec:AccelGibbs}
Any acceleration scheme devised for a stationary linear solver is
a candidate for accelerating convergence of a Gibbs sampler.  For example, consider the second order stochastic iteration
    \be
    \label{ChebySampler}
    \mathbf{y}^{(k+1)}=(1-\alpha_k)\mathbf{y}^{(k-1)} + \alpha_k\f{y}^{(k)} + \alpha_k\tau_k \M^{-1}(\f{c}^{(k)}-\f{Ay}^{(k)})
    \ee
analogous to the linear solver in \eqref{AccScheme} but now the vector $\mathbf{b}$ has been replaced by a random vector $\f{c}^{(k)}$.
The equivalence between polynomial accelerated linear solvers and polynomial accelerated samplers is made clear in the next three theorems.

\begin{theorem}
\label{thm:cheby} Let $\f{A}$ be SPD and $\mathbf{A}=\M -\mathbf{N}$ be a symmetric
splitting.
Consider a set of independent noise vectors $\{\f{c}^{(k)}\}$ with moments
    $$\E(\f{c}^{(k)})=\bnu\mbox{~~and~~}\Var(\f{c}^{(k)})=a_k\M + b_k\mathbf{N}$$
such that
$a_k:= \frac{2-\tau_k}{\tau_k} + \left(b_k - 1\right)\left(\frac{1}{\tau_k} + \frac{1}{\kappa_k} - 1\right)$,
$b_k:= \frac{2(1-\alpha_k)}{\alpha_k}\left(\frac{\kappa_k}{\tau_k}\right) + 1$, $\kappa_{k+1}:=\alpha_k \tau_k + (1-\alpha_k)\kappa_k$, and $\kappa_1=\tau_0$.
If the polynomial accelerated linear solver \eqref{AccScheme} converges to
$\mathbf{A}^{-1}\mathbf{b}$ with a set of parameters $\{\{\alpha_k\}$, $\{\tau_k\}\}$ that are independent of $\{\x^{(k)}\}$, then the polynomial accelerated stochastic iteration \eqref{ChebySampler}
converges in distribution to a distribution $\Pi$ with mean $\A^{-1}\bnu$ and covariance matrix $\A^{-1}$.  Furthermore, if the $\{\f{c}^{(k)}\}$ are normal, then
    $$\y^{(k)}\overset{\mathcal{D}}{\to}\Norm(\bmu=\A^{-1}\bnu,\A^{-1}).$$
\end{theorem}

\proof Appendix~\ref{sect:ap:cheby}. $\Box$

Given a second order linear solver \eqref{AccScheme} that converges, Theorem \ref{thm:cheby} makes clear how to construct a second order sampler \eqref{ChebySampler} that is guaranteed to converge.  The next Corollary shows that the polynomial $P_k$ that acts on the linear solver error $\x^{(k)} - \f{A}^{-1}\f{b}$ is the same polynomial that acts on the errors in the first and second moments of the sampler, $\E(\f{y}^{(k)}) - \A^{-1}\bnu$ and $\Var(\f{y}^{(k)}) - \f{A}^{-1}$ respectively.  In other words, the convergence factors for a polynomial
accelerated solver and sampler are the same.


\begin{corollary}
\label{cor:accmoments}
 Suppose that the polynomial accelerated linear solver \eqref{AccScheme} converges with asymptotic convergence factor $\sigma=\left(\lim_{k\to\infty}\max_{\lambda}|P_k(\lambda)|\right)^{1/k}$, where $P_k$ is the $k^{th}$ order polynomial recursively generated by \eqref{RecursionErrCheby}.  Then under the conditions of Theorem \ref{thm:cheby},
    $$\E\left(\mathbf{y}^{(k+1)}\right) = P_k(\M^{-1}\A)\left(\E(\f{y}^{(0)})-\A^{-1}\bnu\right)\to\A^{-1}\bnu$$
with asymptotic convergence factor $\sigma$, and
    $$\Var\left(\mathbf{y}^{(k)}\right)=\f{A}^{-1} - P_k(\M^{-1}\A)\left(\f{A}^{-1}-\f{Var(y}^{(0)})\right)(P_k(\M^{-1}\A))^T \to\f{A}^{-1}$$
with asymptotic convergence factor $\sigma^2$.
\end{corollary}
\proof Appendix~\ref{sect:ap:cheby}. $\Box$

Corollary \ref{cor:accmoments} allows a direct comparison of the convergence factor for a polynomial
accelerated sampler ($\sigma$, or $\sigma^2$ if $\mu=\f{0}$) to the convergence factor given previously for the
corresponding un-accelerated stationary sampler ($\varrho(\M^{-1}\N)$, or $\varrho(\M^{-1}\N)^2$ if $\mu=\f{0}$).  In particular, given a second order linear solver with accelerated convergence compared to the corresponding stationary iteration, the corollary guarantees that the second order Gibbs sampler \eqref{ChebySampler} will converge faster than the stationary Gibbs sampler \eqref{eq:sampler}.


Just as Chebyshev polynomials are guaranteed to accelerate linear solvers, Corollary \ref{cor:accmoments} assures that Chebyshev polynomials can also accelerate a Gibbs sampler.  Using Theorem \ref{thm:cheby}, we derived the Chebyshev accelerated SSOR sampler \citep{FoxParkerSISC2014} by iteratively updating parameters via \eqref{ChebyAccParams} and then generating a sampler via \eqref{ChebySampler}.   Explicit implementation details of the Chebyshev accelerated sampler are provided in the supplementary materials.   The polynomial accelerated sampler is implemented at a negligible increase in cost of ${\mathcal O}(n)$ flops per iteration over the cost ($4n^2$ flops) of the SSOR sampler (Algorithm \ref{alg:SSORsampler}).
The asymptotic convergence factor is given by the next Corollary, which follows from Corollary~\ref{cor:accmoments} and equation~\eqref{ChebyAccFactor}.

\begin{corollary}
\label{cor:Chebyaccmoments}
If the Chebyshev accelerated linear solver converges, then the mean $\E(\f{y}^{(k)})$ of the corresponding  Chebyshev accelerated stochastic iteration \eqref{ChebySampler} converges to $\bmu=\A^{-1}\bnu$ with asymptotic convergence factor $\left(\frac{1-\sqrt{\lambda_{\min}/\lambda_{\max}}}{1+\sqrt{\lambda_{\min}/\lambda_{\max}}}\right)$ and the covariance matrix $\Var(\f{y}^{(k)})$ converges to $\f{A}^{-1}$ with asymptotic convergence factor $\left(\frac{1-\sqrt{\lambda_{\min}/\lambda_{\max}}}{1+\sqrt{\lambda_{\min}/\lambda_{\max}}}\right)^2$.
\end{corollary}

Corollary \ref{cor:Chebyaccmoments} and \eqref{ChebyPolyOptimal} show that a Chebyshev accelerated normal sampler is guaranteed to converge faster than any other acceleration scheme that has the parameters $\{\{\tau_k, \alpha_k\}\}$ independent of the iterates $\{\y^{(k)}\}$.
 This result also shows that the preconditioning ideas presented in section~\ref{sec:poly} to reduce ${\rm cond}(\f{M}^{-1}\f{A})=\lambda_{\max}/\lambda_{\min}$ can also be used to speed up Chebyshev accelerated samplers.  We do not investigate such preconditioning here.

Corollary \ref{cor:Chebyaccmoments} and equation~\eqref{eq:secondordersolveriterations}
suggest that, for any $\varepsilon >0$, after $k^*$ iterations the Chebyshev error reduction for the mean is smaller than $\varepsilon$.   But even sooner, after $k^{**} = k^*/2$
iterations, the Chebyshev error reduction for the variance is predicted to be smaller than $\varepsilon$ \citep{FoxParkerSISC2014}.




\begin{figure}
\centerline{
\begin{tabular}{{p{0.4\textwidth} p{0.6\textwidth}}}
  \vspace{5pt} \includegraphics[width=0.4\textwidth]{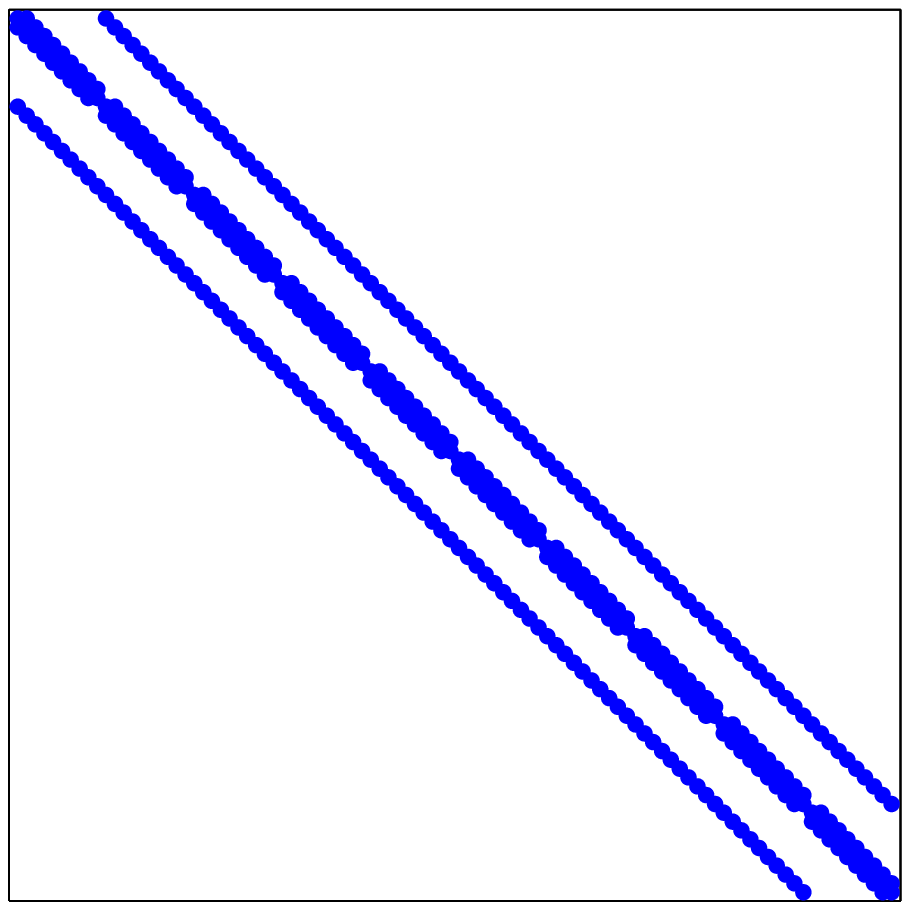} &
  \vspace{0pt} \includegraphics[width=0.6\textwidth]{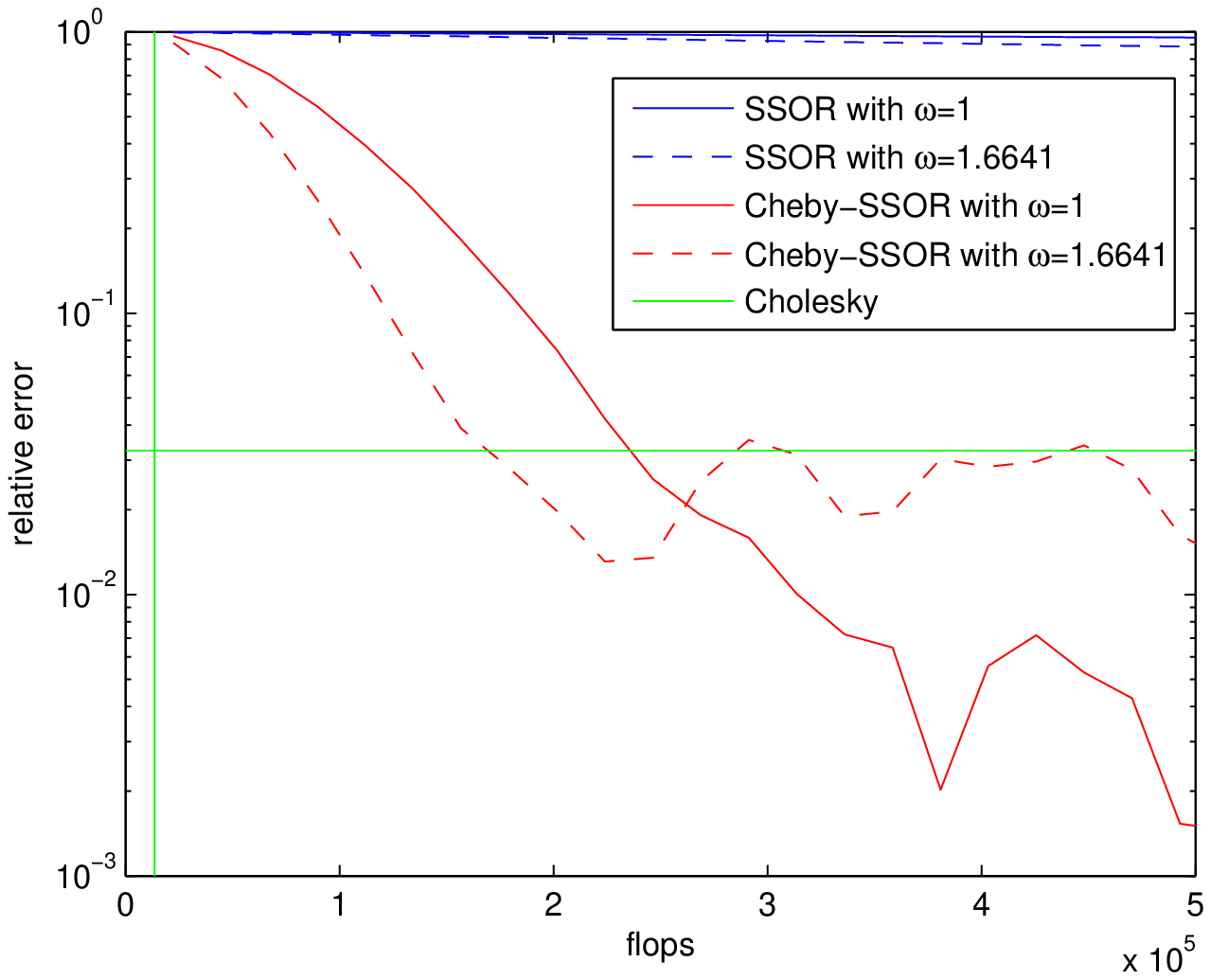}
\end{tabular}
}
\caption{{\small Left panel: Location of non-zero elements in the $100\times 100$ precision matrix $\A$.   Right panel: Relative error in covariance $||\mathbf{A}^{-1}-\f{S}^{(k)}_y||_2/||\f{A}^{-1}||_2$ versus number of floating point operations (flops) for a sampler implemented with SSOR and $\omega=1$, SSOR with optimal relaxation $\omega=1.6641$, and SSOR with Chebyshev acceleration. Also shown is the relative error and flop count for a sample drawn using Cholesky factoring.
}}
  \label{fig:Lap2d}
\end{figure}

\section{Computed Examples}
\label{sect:num}

The iterative sampling algorithms we have investigated are designed for problems
where \emph{operating} by the precision matrix is cheap. A common such case is
when the precision matrix is sparse, as occurs when modeling a GMRF with a
local neighbourhood structure. Then, typically, the precision matrix has
$\mathcal{O}(n)$ non-zero elements, so direct matrix-vector multiplication has
$\mathcal{O}(n)$ cost. We give two examples of sampling using sparse
precision matrices: first, we present a small $n=100$ example where complete
diagnostics can be computed for evaluating the quality of convergence; and second, we present a $n=10^6$ Bayesian linear inverse problem that demonstrates computational feasibility for large problems.  The samplers are initialized with $\f{y}^{(0)}=\f{0}$ in both examples.

\subsection{A $10\times 10$ lattice example ($n=100$)}

A {\em first order locally linear} sparse precision matrix $\A$, considered by~\cite{Higdon,RueHeld}, is
   \bes
   [\A]_{ij}= 10^{-4}\delta_{ij} + \left\{\ba{cc}n_i&~~~{\rm if}~~i=j\\-1&~~~{\rm if} ~~i\ne j ~~{\rm and}~~ ||s_i-s_j||_2\le 1
    \\0& ~~ {\rm otherwise}\ea.\right.
    \ees
The discrete points $\{s_i\}$ are on a regular $10\times 10$ lattice ($n=100$)
over the two dimensional domain ${\cal S}=[1,10]\times[1,10]$.  Thus $\A$ is $100 \times 100$, $||\A||_2=7.8$ and $||\bSigma=\A^{-1}||_2=10^4$.   The sparsity of $\A$ is shown in the left panel of Figure \ref{fig:Lap2d}.   The scalar $n_i$ is the number of points neighbouring $s_i$, i.e., with distance $1$ from $s_i$.  Although $n^2=10^4$, the number of non-zero elements of $\A$ is $\mathcal{O}(n)$ ($460$ in this example).  Since the bandwidth of $\A$ is $\mathcal{O}(n^{1/2})$, a Cholesky factorization costs $\mathcal{O} (n^2)$ flops \citep{Rue2} and each iteration of an iterative method costs $\mathcal{O}(n)$ flops.

To provide a comparison between linear solvers and samplers, we solved the system $\f{Ax}=\mathbf{b}$ using linear solvers with different matrix splittings (Table \ref{tab:splittings}), where $\f{b}$ is fixed and non-zero, all initialized with $\mathbf{x}^{(0)}=\f{0}$.   The results are given in Table~\ref{tab:opt}.   The Richardson method does not converge (DNC) since the spectral radius of the iteration operator is greater than 1.  The SOR iteration was run at the optimal relaxation parameter value of $\omega=1.9852$. SSOR was run at its optimal value of $\omega=1.6641$.
Chebyshev accelerated SSOR (Cheby-SSOR), CG accelerated SSOR (CG-SSOR) (both run with $\omega=1.6641$) and CG utilize a different implicit operator for each iteration, and so the spectral radius given in these cases is the geometric mean spectral radius of these operators (estimated using \eqref{eq:error}).
Even for this small example, Chebyshev acceleration reduces the computational effort required for convergence by about two orders of magnitude, while CG acceleration reduces work by nearly two more orders of magnitude.

\begin{table}
\caption{\label{tab:opt}The number of iterations and the
total number of floating point operations performed by some common stationary and accelerated linear solvers,
and the Cholesky factorization, used to solve $\mathbf{Ax=b}$ for fixed non-zero $\mathbf{b}$. Each solver was run
until the residual became sufficiently small,
$||\mathbf{b}-\A\x^{(k)}||_2<10^{-8}$.  Details in
section~\ref{sect:num}. }
\centering
\fbox{%
\begin{tabular}{|r|c|c|c|l|}
  solver &  $\omega$ & $\varrho(\mathbf{M}^{-1}\mathbf{N})$ & number of
iterations & flops\\
  \hline
   Richardson   &  1        & 6.8           &  DNC              & --
     \\
   Jacobi       &  --       & .999972       & $4.01 \times 10^5$& $5.69\times
10^7$     \\
   Gauss-Seidel & --         & .999944       & $2.44 \times 10^5$& $4.34\times
10^8$             \\
   SSOR         & 1.6641    & .999724       & $6.7\times 10^4$  & $2.39\times
10^8$    \\
   SOR          & 1.9852    & .985210       & 1655              & $2.95\times
10^6$  \\
   Cheby-SSOR   & 1         & .9786         & 958               & $3.41\times
10^6$    \\
   Cheby-SSOR   & 1.6641    & .9673         & 622               & $2.21\times
10^6$    \\
   CG           & --        & .6375         & 48                & $9.22\times
10^4$   \\
   CG-SSOR      & 1.6641    & .4471         & 29                & $6.66\times
10^4$   \\
   Cholesky     &  --       & --            & --                & $1.35\times
10^4$
\end{tabular}}
\end{table}

We investigated the following
Gibbs samplers: SOR, SSOR, and the Chebyshev accelerated SSOR. These samplers are guaranteed to converge since the corresponding solver converges (Theorem~\ref{thm:StatConvSampConv}).  Since the convergence factor for a sampler is equal to the convergence factor for the corresponding solver (Corollaries~\ref{cor:StatConvSampConv} and \ref{cor:accmoments}) then Gibbs samplers  implemented with any of the matrix splittings in Table \ref{tab:splittings} exhibit the same convergence behavior as shown for the linear solvers in Table \ref{tab:opt}. Convergence of the sample covariance $\f{S}^{(k)}_y
\approx \f{Var(y}^{(k)})\to\mathbf{A}^{-1}$, calculated using $10^4$ samples, is shown in the right panel of
Figure \ref{fig:Lap2d} that displays the relative error $||\mathbf{A}^{-1}-\f{S}^{(k)}_y||_2/||\f{A}^{-1}||_2$ as a function of the flop count.  Each sampler iteration costs
about $2.24\times 10^3$ flops.  This performance is compared to samples
constructed by a Cholesky factorization, which cost $1.34\times 10^4$ flops (depicted as the green vertical line in the right panel of Figure \ref{fig:Lap2d}).
Since the sample means were uniformly close to zero, error in the mean is not shown.

The benchmark for evaluation of the of the convergence of the iterative samplers in finite precision is the Cholesky factorization, its relative error is depicted as the green horizontal line in the right panel of Figure \ref{fig:Lap2d}.
For this example, the iterative samplers produce better samples than a Cholesky sampler since the iterative sample covariances become more precise 
with more computing time.

The geometric convergence in distribution of the un-accelerated SSOR samples $\mathbf{y}^{(k)}$ to $\Norm(\f{0},\mathbf{A}^{-1})$  is clear in Figure \ref{fig:Lap2d}, and even after $5\times 10^5$ flops, convergence in distribution has not been attained.   This is not surprising based on the large number of iterations ($\mathcal O(n^2)$) necessary for the same stationary method to converge to a solution of $\f{Ax}=\mathbf{b}$ (see Table \ref{tab:opt}).   The accelerated convergence of the Chebyshev polynomial samplers, suggested by the fast convergence of the corresponding linear solvers depicted in Table \ref{tab:opt}, is also evident in Figure \ref{fig:Lap2d}, with convergence after
$1.70\times 10^5$ flops (76 iterations)      
for the Cheby-SSOR sampler with optimal relaxation parameter $\omega=1.6641$, and the somewhat slower convergence at
$2.37\times 10^5$ flops (106 iterations)     
when $\omega=1$.

\subsection{A $100\times 100\times 100$ ($n=10^6$) linear inverse problem in biofilm imaging}
We now perform accelerated sampling from a GMRF in 3-dimensions, as a stylized example of estimating a voxel image of a biofilm from confocal scanning laser microscope (CSLM) data~\citep{Lewandowski2014}. This large example illustrates the feasibility of Chebyshev accelerated sampling in large problems for which sampling by Cholesky factorization of the precision matrix is too computationally and memory intensive to be performed on a standard desktop computer. 

We consider the problem of reconstructing a $100\times 100\times 100$ voxel image $\x$ of a bacterial biofilm, i.e., a community of bacteria aggregated together as \emph{slime}, given a subsampled $100\times 100\times 10$ CSLM data set $\y$. For this exercise, we synthesized a `true' image $\x_\text{t}$ of a $90\mu$m tall ellipsoidal column of biofilm attached to a surface, taking value $10$ inside the biofilm column, and $0$ outside, in arbitrary units.  Similar geometry has been observed experimentally for {\it Pseudomonas aeruginosa} biofilms~\citep{SwoggerPitts}, and is also predicted by mathematical models of biofilm growth \citep{Klapper2005}. CSLM captures a set of planar `images' at different distances from the bottom of the biofilm where it is attached to a surface.  In nature biofilms attach to any surface over which water flows, e.g., human teeth and creek bottoms. Each horizontal planar image in this example is $100 \times 100$ pixels; the distance between pixels in each plane is typically about 1 $\mu$m, with the exact spatial resolution set by the microscope user.  The vertical distance between planar slices in a CSLM image is typically an order of magnitude larger than the horizontal distance between pixels; for this example, the vertical distance between CSLM planes is 10$\mu$m.

\begin{figure}
\centerline{
\begin{tabular}{cc}
\includegraphics[height=2.4in,width=3.4in]{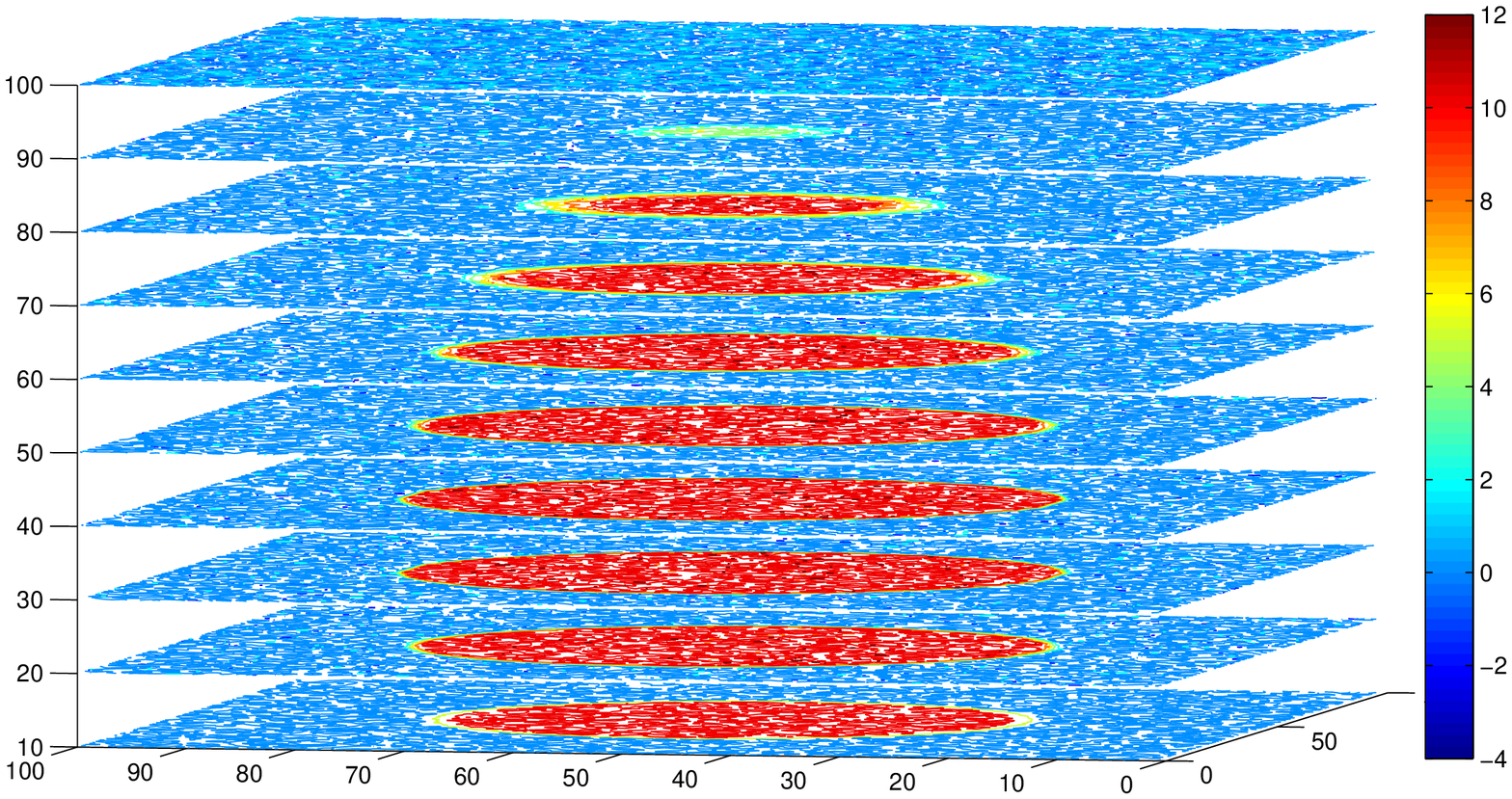}&
\includegraphics[height=2.4in]{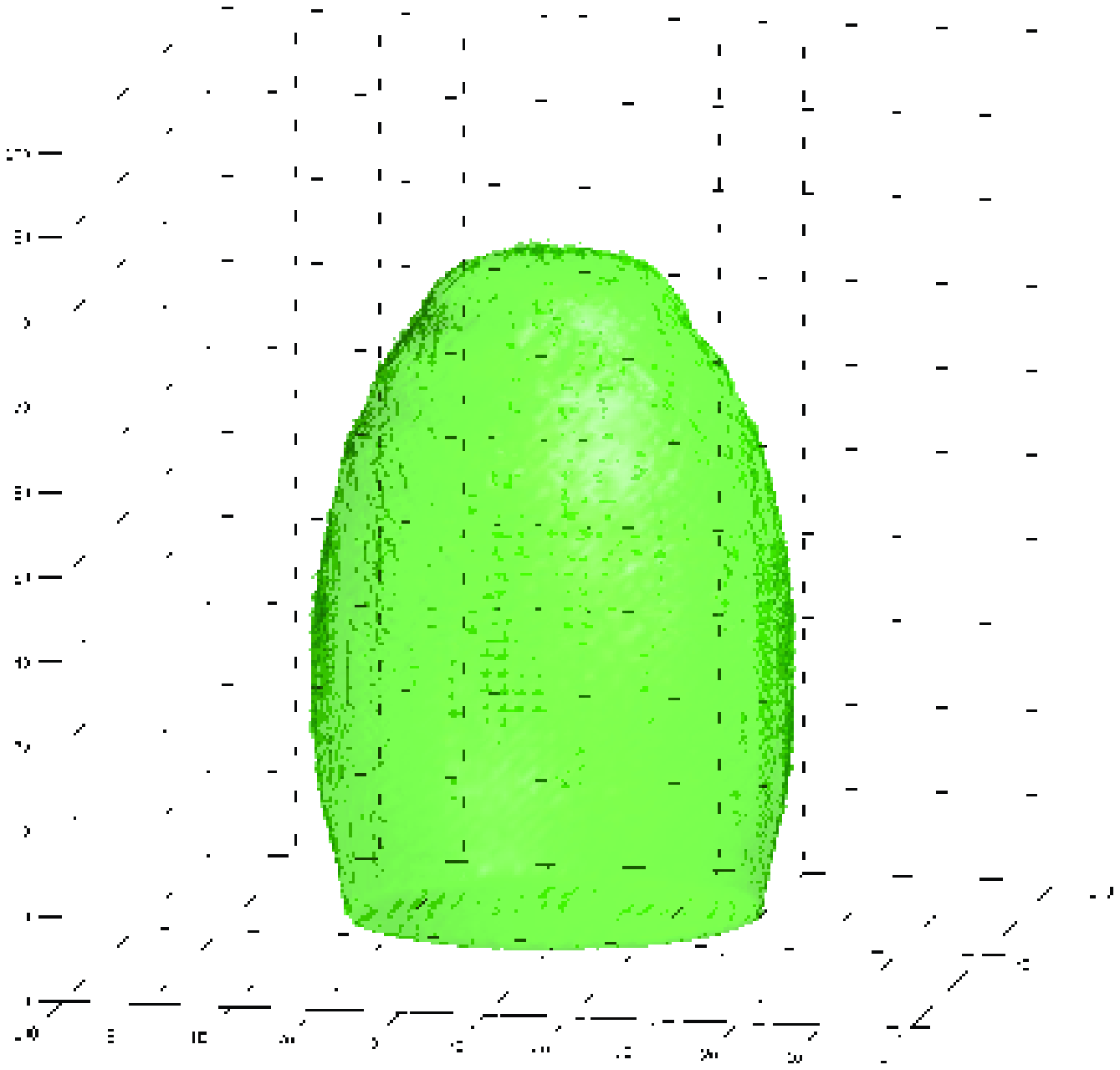}
\end{tabular}}
\caption{{\small The left panel depicts a $100\times 100\times 10$ pixelated confocal scanning microscope image, $\f{y}$ of a simulated ellipsoidal column of a bacterial biofilm; the distance between horizontal pixels is 1$\mu$m, the distance between vertical pixels is 10$\mu$m.  The right panel shows a surface rendering of a sample from the $n=10^6$ dimensional multivariate normal posterior distribution conditioned on hyperparameters.}}
  \label{fig:Lap3d}
\end{figure}

Given the `true' image $\x_\text{t}$, we generated synthetic $100\times 100\times 10$ CSLM data by
    $$\f{y}=\f{F}\x_\text{t} + \epsilon$$
where the $10^5\times 10^6$ matrix $\f{F}$ arithmetically averages over 10 pixels in the vertical dimension of $\x$, to approximate the point spread function (PSF) of CSLM~\citep{ShepardShotton}, and $\epsilon\sim \Norm(\f{0},\bP^{-1}=\f{I}$). The data is displayed in the left panel of Figure~\ref{fig:Lap3d} as layers of pixels, or `slices', located at the centre of sensitivity of the CSLM, i.e. the centre of the PSF.  Thus, the likelihood we consider is $\pi(\f{y}|\x)=\Norm(\f{F}\x,\bP^{-1})$.

To encapsulate prior knowledge that the bacteria in the biofilm aggregate together we model $\x$ by the GMRF $\x\sim N(\f{0},\Q_R^{-1})$ where the precision matrix $\Q_R$ models local smoothness of the density of the biofilm and background. We construct the matrix $\Q_R$ as a sparse inverse of the dense covariance matrix corresponding to the exponential covariance function. This construction uses the relationship between stationary Gaussian random fields
and partial differential equations (PDEs) that was noted by \cite{Whittle1954} for the
Mat\'{e}rn (or Whittle-Mat\'{e}rn \citep{GuttorpGneiting2005}) class of
covariance functions, that was also exploited by
\cite{cui2011bayesian} and \cite{LindgrenRueLindstromJRSSB}. Rather
than stating the PDE, we find it more convenient to work with the equivalent
variational form, in this case (the square of)%
\[
\mathcal Q\left(  x\right)  =\int_{\mathcal{D}}\left(  \frac{R}{4}\left\vert \nabla
x\right\vert ^{2}+\frac{1}{4R}x^{2}\right)  dv+\int_{\partial\mathcal{D}}\frac{x^{2}}{2}ds,
\]
where $x$ is a continuous stochastic field, $dv$ is the volume element in the domain $\mathcal{D}$ and $ds$ is the surface element on the boundary $\partial\mathcal{D}$. This form has Euler-Lagrange equations being the Helmholtz operator with (local) Robin
boundary conditions $x+R\frac{\partial  x}{\partial n}=0$ on
$\partial\mathcal{D}$, induced by the $\frac{x^{2}}{2}$ term. In our example we apply the Hessian of this form twice, which can be
thought of as squaring the Helmholtz operator. When the quadratic form is written in the operator form ${\mathcal Q}\left(  x\right)
=x^{T}Hx$, where $H$ is the Hessian, the resulting Gaussian random field has density%
\be
\label{xprior}
\pi\left(  x\right)  \propto\exp\left\{  -x^{T}H^{2}x\right\}  .
\ee
We chose this operator because the discretized precision matrix is sparse, while the
covariance function (after scaling) is close to $\exp\left\{  -r/R\right\}$, having length-scale $R$.

\begin{figure}
\centerline{
\begin{tabular}{cc}
\includegraphics[width=0.4\textwidth]{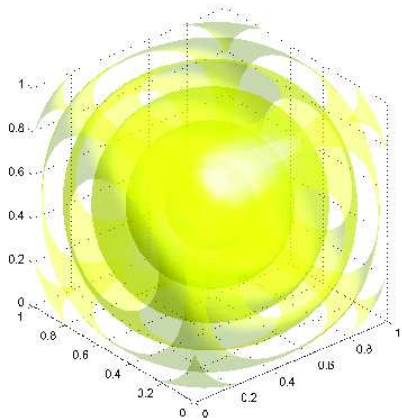}&
\end{tabular}}
\caption{\small Contours of the effective covariance function centred on the cubic domain, logarithmically spaced in value. }
  \label{fig:CovSurf}
\end{figure}

The GMRF over the discrete field $\x$ is then defined using FEM (finite element method) discretization; we used cubic-elements between nodes at voxel centres in the cubic domain, and tri-linear interpolation from nodal values within each element.
To verify this construction we show in Figure~\ref{fig:CovSurf} contours of the resulting covariance function, between the pixel at the centre of the normalised cubic domain $\left[0,1\right]^3$ and all other pixels, for length scale $R=1/4$.  The contours are logarithmically spaced in value, hence the evenly spaced spherical contours show that the covariance indeed has exponential dependence with distance. The contours look correct at the boundaries, indicating that the local Robin boundary conditions\footnote{Local boundary conditions are approximate but preserve sparseness. The exact boundary conditions are given by the boundary integral equation for the exterior Helmholtz operator, resulting in a dense block in $H$ that is inconvenient for computation~\citep{NeumayerPhD}.} give the desired covariance function throughout the domain. In contrast, Dirichlet conditions would make the cubic boundary a contour, while Neumann conditions as used by~ \cite{LindgrenRueLindstromJRSSB} would make contours perpendicular to the cubic boundary; neither of those pure boundary conditions produce the desired covariance function.

In the deterministic setting, this image recovery problem is an example of a linear inverse problem. In the Bayesian setting, we may write the hierarchical model in the general form
\begin{eqnarray}
\y|\x,\btheta & \sim & \Norm\left(\F\x,\bP_{\btheta}^{-1}\right)\label{eq:yxtheta}\\
\x|\btheta & \sim & \Norm\left(\bmu,\Q_{\btheta}^{-1}\right)\label{eq:xtheta}\\
\btheta & \sim & \pi\left(\btheta\right)\label{eq:theta}
\end{eqnarray}
where $\btheta$ is a vector of hyperparameters. This stochastic model
occurs in many settings~\citep[see, e.g.,][]{SimpsonLindgrenRue2012,RueHeld} with $\y$ being \emph{observed data}, $\x$ is a \emph{latent field},
and $\btheta$ is a vector of \emph{hyperparameters} that parameterize the precision matrices $\bP$ and $\Q$. The (hyper)prior $\pi\left(\btheta\right)$ models uncertainty in covariance of the two random fields.

There are several options for performing sample-based inference on the model \eqref{eq:yxtheta}, \eqref{eq:xtheta}, \eqref{eq:theta}. Most direct is forming the posterior distribution $\pi\left(\x,\btheta|\y\right)$ via Bayes' rule and implementing Markov chain Monte Carlo (MCMC) sampling, typically employing Metropolis-Hastings dynamics with a random walk proposal on $\x$ and $\btheta$.  Such an algorithm can be very slow due to high correlations within the latent field $\x$, and between the latent field and hyperparameters $\btheta$. More efficient algorithms block the latent field, noting that the distribution over $\x$ given everything else is a multivariate normal, and hence can be sampled efficiently as we have discussed in this paper. \cite{Higdon} and \cite{BardsleyRTO} utilized this structure, along with conjugate hyperpriors on the components of $\btheta$, to demonstrate a Gibbs sampler that cycled through sampling  from the conditional distributions for $\x$ and components of $\btheta$. When the normalizing constant for $\pi\left(\x,\btheta|\y\right)$ is available, up to a multiplicative constant independent of state, a more efficient algorithm is the \emph{one block} algorithm \citep[][section 4.1.2]{RueHeld} in which a candidate $\btheta'$ is drawn from a random walk proposal, then a draw $\x'\sim \pi(\x'|\y,\btheta')$, with the joint proposal $\left(\btheta',\x'\right)$  accepted with the standard Metropolis-Hastings probability. The resulting transition kernel in $\btheta$ is in detailed balance with the distribution over $\btheta|\y$, and hence can improve efficiency dramatically. A further improvement can be to perform MCMC directly on $\pi\left(\btheta|\y \right)$ as indicated by \cite{SimpsonLindgrenRue2012}, with subsequent independent sampling $\x\sim\pi\left(\x|\y,\btheta \right)$ to facilitate Monte Carlo evaluation of statistics. In each of these schemes, computational cost is dominated by the cost of drawing samples from the large multivariate normal $\x\sim\pi\left(\x|\y,\btheta \right)$. We now demonstrate that sampling step for this synthetic example.

In our example, the distribution over the $100\times 100\times 100$ image $\x$, conditioned on everything else, is the multivariate normal
    \be
    \label{eqn:BIGnormal}
    \pi(\x|\f{y},\theta=R)&=&\Norm(\x;\bmu=\f{A}^{-1}\f{F}^T\bP\y,\bSigma=\f{A}^{-1})
    \ee
with precision matrix $\f{A}=\f{F}^T\bP\F + \Q_R$ (cf. \cite{Calvetti,Higdon}). For this calculation we used the same covariance matrix as shown above, so $R=1/4$ in units of the width of the domain, though for sample-based inference one would use samples from the distribution over $R|\x,\y$.
The right panel of Figure~\ref{fig:Lap3d} depicts a reconstructed surface derived from a sample from the conditional distribution in \eqref{eqn:BIGnormal} using the Chebyshev polynomial accelerated SSOR sampler.  
The sampler was initialized with the precision matrix $\A$, $\E(\f{c}^{(k)})=\f{F}^T\bP\f{y}$ for all $k$, and relaxation parameter $\omega=1$. The contour is at value $6$, after smoothing over $3\times3\times3$ voxels, displaying a sample surface that separates regions for which the average over $3\times3\times3$ voxel blocks is less than $6$ (outside surface) and greater than $6$ (inside surface). As can be seen, the surface makes an informative reconstruction of the ellipsoidal phantom.

Using CG,
estimates of the extreme eigenvalue of $\M_{\text{SSOR}}^{-1}\A$ were $\hat \lambda_{\min} = 4.38 \times 10^{-6}$ and $\hat \lambda_{\max} = 1 - 1.36 \times 10^{-8}$.  By Corollary 6, the asymptotic convergence factors for the Chebyshev sampler are $\sigma\approx 0.9958$ for the mean and $\sigma^2 = .9917$ for the covariance matrix.   Using this information, equation 
\eqref{eq:secondordersolveriterations}
predicts the number of sampler iterations until convergence.   After $k^*=4566$ iterations of the Chebyshev accelerated sampler, it is predicted that the mean error is reduced by $\varepsilon=10^{-8}$; that is
    $$||\bmu - \E(\y^{(k^*)})||_2 \approx 10^{-8}||\bmu - \E(\y^{(0)})||_2.$$
But even sooner, after only $k^{**}=k^*/2=2283$ iterations,  it is predicted that the covariance error is
      $$||\A^{-1} - \Var(\y^{(k^{**})})|| \approx 10^{-8}||\A^{-1} - \Var(\y^{(0)})||.$$

Contrast these Chebyshev polynomial convergence results to the performance of the non-accelerated stationary SSOR sampler that has convergence factors $\varrho(\M^{-1}\N) \approx 1 - \hat \lambda_{\min} = 1 - 4.38 \times 10^{-6}$ for the mean error, and $\rho(\M^{-1}\N)^2 = 1 - 8.76\times 10^{-6}$ for covariance error.  These convergence factors suggest that after running the non-accelerated SSOR Gibbs sampler for only $4566$ iterations, the covariance error will be reduced to only $\varrho(\M^{-1}\N)^{2\cdot 4566}\approx 0.96$ of the original error; $1.9\times 10^6$ iterations are required for a $10^{-8}$ reduction.

The cost difference between the Cholesky factorization and an iterative sampler in this example is dramatic. After finding a machine with the necessary $n^2$ memory requirements, the Cholesky factorization would cost about $b^2n=10^{16}$ flops (since the bandwidth of the precision matrix $\A$ is about $b=10^5$).  Since the number of non-zero elements of $\A$ is $3.3\times 10^8$, an iterative sampler costs about $6.6\times 10^8$ flops per iteration, much less than $n^2$.    The sample in Figure \ref{fig:Lap3d} 
was generated by
$k_{\max}=5\times 10^3$ iterations of the Chebyshev accelerated SSOR sampler, at a total cost of $3.3 \times 10^{12}$ flops, which is about $10^4$ times faster than Cholesky factoring.

\section{Discussion}
This work began, in part, with a curiosity about 
the convergence of the sequence of covariance matrices in Gibbs sampling applied to multivariate normal distributions, as studied by~\cite{LiuWongKong95}. Convergence of that sequence indicates that the algorithm is implicitly implementing some factorization of the target covariance or precision matrix. Which one?

The answer was given by \cite{GoodmanSokalMGMC}, \cite{AmitGrenander}, \cite{Barone}, and  \cite{GalliGao}, that the standard component-sweep Gibbs sampler corresponds to the classical Gauss-Seidel iterative method. %
That result is given in section \ref{sec:SampNormSplitting}, generalized to arbitrary matrix splittings, showing that
any matrix splitting used to generate a deterministic relaxation also induces a stochastic relaxation that is a generalized Gibbs sampler; the linear iterative relaxation and the stochastic relaxation share exactly the same iteration operator, conditions for convergence, and convergence factor, which may be summarized by noting that they share exactly the same error polynomial. 

Equivalence of error polynomials is important because they are the central object in 
designing accelerated solvers including the multigrid, Krylov space, and parallel algorithms. We demonstrated that equivalence explicitly for polynomial acceleration, the basic non-stationary acceleration scheme for linear solvers, showing that this control of the error polynomial can be applied to Gibbs sampling from normal distributions. It follows that, just as for linear solvers, Chebyshev-polynomial accelerated samplers have a smaller average asymptotic convergence factor than their un-accelerated stationary counterparts. 

The equivalences noted above are strictly limited to the case of normal target distributions. We are also concerned with continuous non-normal target distributions and whether acceleration of the normal case can usefully inform acceleration of sampling from non-normal distributions. Convergence of the unaccelerated, stationary, iteration applied to bounded perturbations of a normal distribution was established by~\cite{Amit1991}, though carrying over convergence rates proved more problematic.

There are several possibilities for extending the acceleration techniques to non-normal distributions. 
A straightforward generalization is to apply Gibbs sampling to the non-normal target, assuming the required conditional distributions are easy to sample from, though using the directions determined by the accelerated algorithm. Simply applying the accelerated algorithm to the non-normal distribution does not lead to optimal acceleration, as demonstrated by~\cite{GoodmanSokalMGMC}. 

A second route, that looks more promising to us, is to exploit the connection between Gibbs samplers and linear iterative methods that are often viewed as \emph{local} solvers for non-linear problems, or equivalently, optimizers for local quadratic approximations to non-quadratic functions. Since a local quadratic approximation to $\log\pi$ is a local Gaussian approximation to $\pi$, the iterations developed here may be used to target this local approximation and hence provide local proposals in an MCMC. We imagine an algorithm along the lines of the trust-region methods from optimization in which the local quadratic (Gaussian) approximation is trusted up to some distance from the current state, implemented via a distance penalty. One or more steps of the iterative sampler would act as a proposal to a Metropolis-Hastings accept/reject step that ensures the correct target distribution. Metropolis adjusted Langevin (MALA) and hybrid Monte Carlo (HMC) turn out to be examples of this scheme~\citep{NortonFox}, as is the algorithm presented by~\cite{GreenHan}. This naturally raises the question of whether acceleration of the local iteration can accelerate the Metropolised algorithm. This remains a topic for ongoing research.

\section*{Acknowledgments} This work was partially funded by the New Zealand
Institute for Mathematics and its Applications (NZIMA) thematic
programme on Analysis, Applications and Inverse Problems in PDEs, and Marsden contract UOO1015.

\appendix
\section{Appendix}

\subsection{Stationary sampler convergence (Proof of Theorem \ref{thm:momentconv} and Corollary \ref{cor:StatConvSampConv})}
\label{sect:ap:VarB}
First, the theorem and corollary are established for the mean.
Since $\A=\M-\N$ is a convergent splitting, then
\eqref{eq:sampler} and Theorem \ref{thm:StatConvSampConv} show that $E(\f{c}^{(k)})=\bnu$ if and only if $E(\y^{(k)})\to\A^{-1}\bnu$
with the same convergence factor as for the linear solver.  To establish convergence of the variance, let $\f{G=M}^{-1}\f{N}$ in \eqref{eq:sampler}, then
    $\f{y}^{(k)}=\f{G}^k\f{y}^{(0)} + \sum_{i=0}^{k-1}\f{G}^i(\f{M}^{-1}\f{c}^{(k-1-i)}).$
This equation and the
independence of $\{\mathbf{c}^{(i)}\}$ show that
    $\f{Var(y}^{(k)}|\mathbf{y}^{(0)}) = \sum_{i=0}^{k-1}\left(\mathbf{G}^i\f{M}^{-1}\Var(\mathbf{c}^i)\f{M}^{-T}
(\mathbf{G}^i)^T\right).$
Theorem \ref{thm:StatConvSampConv}
establishes the existence of a unique limiting distribution
with a non-zero covariance matrix  $\f{\Gamma}$.  Thus, for $\f{y}^{(i)},\f{y}^{(i+1)}\sim \Pi$,
\eqref{eq:sampler} implies
    \be
    \label{varBrelation}
    \f{\Gamma}=\f{G\Gamma}\mathbf{G}^T + \M^{-1}\Var(\mathbf{c}^{(i)})\M^{-T}
    \ee
since $\f{y}^{(i)}$ and $\mathbf{c}^{(i)}$ are independent.
Thus
    $\f{Var(y}^{(k)}|\mathbf{y}^{(0)}) = \mathbf{\Gamma} -\mathbf{G}^k\f{\Gamma}(\mathbf{G}^k)^T,$
and so
    \be
    \f{Var(y}^{(k)}) = \f{\Gamma} -
\f{G}^k\left(\f{\Gamma}-\f{Var(y}^{(0)})\right)(\f{G}^k)^T.\label{InitAcc}
    \ee
That is, $\Var(\y^{(k)})\to\mathbf{\Gamma}$ with convergence factor
    $\varrho(\M^{-1}\mathbf{N})^2$.
To prove that part (b) of the theorem implies part (a), consider the starting
vector $\f{y}^{(0)}\sim \Pi$ with covariance matrix
$\f{\Gamma=A}^{-1}$.  Since $\f{c}^{(k)}$ is independent of
$\f{y}^{(k)}$, the relation~\eqref{varBrelation} shows that
    $\Var(\mathbf{c}^{(k)})=\M
(\mathbf{A}^{-1}-\f{GA}^{-1}\mathbf{G}^T)\M^T=\f{MA}^{-1}\M^T-\f{NA}^{-1}\mathbf
{N}^T$.
Substituting in $\f{N=M-A}$ shows that
$\Var(\mathbf{c}^{(k)})=\f{M}^T+\f{N}$.
To prove that (a) implies (b), consider $\f{y}^{(0)}\sim \Norm(\bmu,\f{A}^{-1})$.   By \eqref{InitAcc},
$\f{\Gamma}-
\Var(\f{y}^{(1)})=\f{G}(\f{\Gamma}-\f{A}^{-1})\f{G}^T$.
Substituting $\Var(\mathbf{c}^{(0)})=\M
(\mathbf{A}^{-1}-\f{GA}^{-1}\mathbf{G}^T)\M^T$ into equation
\eqref{varBrelation} shows $\f{\Gamma}-
\f{A}^{-1}=\f{G}(\f{\Gamma}-\f{A}^{-1})\f{G}^T$.   Thus
$\Var(\f{y}^{(1)})=\f{A}^{-1}$, which shows that
$\Var(\f{y}^{(k)})$ has converged to $\f{A}^{-1}$.  By Theorem
\ref{thm:StatConvSampConv},
$\f{\Gamma = A}^{-1}$.

\subsection{Polynomial accelerated sampler convergence (Proof of Theorem~\ref{thm:cheby} and Corollary \ref{cor:accmoments})}
\label{sect:ap:cheby}
If the polynomial accelerated linear solver \eqref{AccScheme} converges, then
$E(\f{y}^{(k+1)})\to \f{A}^{-1}E(\f{c}^{(k)})=\bmu$.
To determine
$\Var(\f{c}^{(k)})$
rewrite the iteration \eqref{ChebySampler}
as
    $\f{y}^{(k+1)} = (1-\alpha_k) \f{y}^{(k-1)} + \alpha_k\f{ G}^{(k)} \f{y}^{(k)} + \alpha_k \left(\f{ M}^{(k)}\right)^{-1}\f{c}^{(k)}$
where $\f{ M}^{(k)}=\frac{1}{\tau_k}\f{M}$, $\f{ N}^{(k)}=\f{ M}^{(k)}-\f{A}$, and $\f{ G}^{(k)}=\f{I} - \tau_k \left(\f{ M}^{(k)}\right)^{-1}\f{A} = \left(\f{ M}^{(k)}\right)^{-1} \f{ N}^{(k)}$.
First, we will consider $\f{y}^{(i-1)}\sim \Norm(\f{\mu,A}^{-1})$ and then find $\Var(\f{c}^{(k)})$ that will guarantee that $\f{y}^{(i)},\f{y}^{(i+1)}\sim \Norm(\bmu,\f{A}^{-1})$.   Since $\{\f{c}^{(i)}\}$ are independent of $\{\f{y}^{(i)}\}$,
the above equation for $\f{y}^{(k+1)}$ shows that $\Var(\f{c}^{(k)})$ is equal to
    \bes
        \frac{1}{\alpha_k^2}\f{ M}^{(k)} \left((1-(1-\alpha_k)^2)\f{A}^{-1} - 2(1-\alpha_k)\alpha_k(\f{ G}^{(k)}\f{K}^{(k)} + \f{K}^{(k)T}\f{ G}^{(k)T})-\alpha_k^2\f{ G}^{(k)}\f{A}^{-1}\f{ G}^{(k)T}   \right)\f{ M}^{(k)}
    \ees
where $\f{K}^{(k)}:=\f{Cov}(\f{y}^{(k-1)},\f{y}^{(k)})$.  To simplify this expression, we need Lemma \ref{lem:Cheby}, which gives $\f{K}^{(k)}$ explicitly.  Parts (\ref{lem:Cheby1}) and (\ref{lem:Cheby2}) of the lemma show that
    \bes
        \Var(\f{c}^{(k)})=\frac{1}{\alpha_k^2}\f{ M}^{(k)} \left(\alpha_k^2(\f{A}^{-1} - \f{ G}^{(k)}\f{A}^{-1}\f{ G}^{(k)T})
        + 2(1-\alpha_k)\alpha_k(\f{A}^{-1} - \f{ G}_{\kappa}^{(k)}\f{A}^{-1}\f{ G}^{(k)T})   \right)\f{ M}^{(k)}.
    \ees
Part (\ref{lem:Cheby3}) of Lemma \ref{lem:Cheby} shows that $\Var(\f{c}^{(k)})$ has the form specified in the theorem.

\begin{lemma}
\label{lem:Cheby}
For a symmetric splitting $\A=\M-\N$,
\begin{enumerate}
\item \label{lem:Cheby1} $\f{K}^{(k)}$ is symmetric.

\item \label{lem:Cheby2} $\f{K}^{(k)}=\f{G}_{\kappa}^{(k)}\f{A}^{-1}$, where $\f{G}_{\kappa}^{(k)}=\f{I} - \kappa_k \f{ M}^{-1}\f{A}$ and
        $\kappa_{k+1}:=\alpha_k\tau_k + (1-\alpha_k)\kappa_k.$

\item \label{lem:Cheby3} $\f{A}^{-1} - \f{ G}_{\tau}\f{A}^{-1}\f{ G}_\kappa^T = \tau\kappa\f{M}^{-1}\left((1/\tau + 1/\kappa)\f{M} - \f{A}\right)\f{M}^{-1}$.

\end{enumerate}
\end{lemma}

\proof To nail down $\K^{(k)}$, rewrite the Chebyshev iteration \eqref{ChebySampler} as
\begin{equation*}
\f{Y}^{(k+1)}=\left(
\begin{array}
[c]{cc}%
\alpha_{k}\G^{(k)} & \left(  1-\alpha_{k}\right)  \f{I}\\
\f{I} & \f{0}
\end{array}
\right)  \f{Y}^{(k)}+\alpha_{k}\left(
\begin{array}
[c]{c}%
\f{g}^{(k)}\\
\f{0}
\end{array}
\right)  
\end{equation*}
where
$
\f{Y}^{(0)}=\left(
\begin{array}
[c]{c}%
\y^{(0)}\\
\f{0}
\end{array}
\right)  ,\f{Y}^{(k+1)}=\left(
\begin{array}
[c]{c}%
\y^{(k+1)}\\
\y^{(k)}%
\end{array}
\right)\mbox{~~and~~}
\f{g}^{(k)}=\left(\M^{(k)}\right)^{-1}\f{c}^{(k)}.
$
Letting
$
\mathcal{G}^{(k)}=\left(
\begin{array}
[c]{cc}%
\alpha_{k}\G^{(k)} & \left(  1-\alpha_{k}\right)  \f{I}\\
\f{I} & \f{0}
\end{array}
\right)   
$
shows that
\be
\Var\left(  \f{Y}^{(k+1)}\right)     =\mathcal{G}^{(k)}\mathrm{Var}\left(  \f{Y}^{(k)}\right)  \mathcal{G}^{(k)T}%
+\alpha_{k}^{2}\left(
\begin{array}
[c]{cc}%
\Var\left(  \f{g}^{(k)}\right)  & \f{0}\\
\f{0} & \f{0}
\end{array}
\right). \label{eq:ssmvariter}%
\ee
If $\Var\left(  \y^{(0)}\right)  =\A^{-1}$ then
$\Var\left(  \y^{(k)}\right)  =\A^{-1}$ for $k\ge 1$ in which case $\Var\left(  \f{Y}^{(k+1)}\right)$ is 
\be
\left(
\begin{array}
[c]{cc}%
\A^{-1} & \K^{(k+1)}\\
\K^{(k+1)T} & \A^{-1}%
\end{array}
\right)  &=&\mathcal{G}^{(k)}\left(
\begin{array}
[c]{cc}%
\A^{-1} & \K^{(k)}\\
\K^{(k)T} & \A^{-1}%
\end{array}
\right)  \mathcal{G}^{(k)T}+\left(
\begin{array}
[c]{cc}%
\alpha_{k}^2\Var\left(  g^{(k)}\right)  & \f{0}\\
\f{0} & \f{0}
\end{array}
\right).
\label{eq:ssmfixedvar}%
\ee
By definition of $\f{Y}^{(0)}$,
$\K^{(0)}=\f{0}$;
for $k\ge 0$,
\be
\label{Keq}
\K^{(k+1)}=\alpha_{k}\G^{(k)}\A^{-1}+\left(  1-\alpha_{k}\right)  \K^{(k)T}.
\ee
Since $\alpha_0=1$, then 
    $\K^{(1)}=\G^{(0)}\A^{-1}$
which proves parts (\ref{lem:Cheby1}) and (\ref{lem:Cheby2}) of the Lemma for $k=0$ since $\kappa_{1}=\tau_{0}$ and $\G_{\kappa}^{(k)}\A^{-1}$ is symmetric.   Assuming that $\K^{(k)}=\G^{(k)}_\kappa\A^{-1}$ for $k>0$, the recursion in \eqref{Keq} gives
    $\K^{(k+1)} = \left(  \f{I}-\left[  \alpha_{k}\tau_{k}+\left(  1-\alpha_{k}\right)
    \kappa_{k} \right]  \M^{-1}\A\right)  \A^{-1}$
so the expansion and recursion hold for $k+1$, and parts (\ref{lem:Cheby1}) and (\ref{lem:Cheby2}) of the Lemma follow by induction.
Part (c) of the Lemma follows from the equation
\begin{align*}
\A^{-1}-\G_{\tau}\A^{-1}\G_{\kappa}^{\text{T}}  &  =\M_{\tau}^{-1}\left(  \M_{\tau
}\A^{-1}\M_{\kappa}^{\text{T}}\right)  \M_{\kappa}^{-\text{T}}-\M_{\tau}%
^{-1}\N_{\tau}\A^{-1}\left(  \M_{\kappa}^{-1}\N_{\kappa}\right)  ^{\text{T}}. \hspace{0.7in} \Box
\end{align*}

The selection of $\f{Var(c}^{(k)})=a_k\M + b_k\N$ assures that if $\Var\left(  \y^{(0)}\right)  =\A^{-1}$, then
$\Var\left(  \y^{(k)}\right)  =\A^{-1}$ for $k\ge 1$.  Thus, 
subtracting \eqref{eq:ssmfixedvar} from \eqref{eq:ssmvariter}
gives
\begin{equation*}
\Var\left(  \f{Y}^{(k+1)}\right)  -\left(
\begin{array}
[c]{cc}%
\A^{-1} & \K^{(k+1)}\\
\K^{(k+1)T} & \A^{-1}%
\end{array}
\right)  =\mathcal{G}^{(k)}\left(  \Var\left(  \f{Y}^{(k)}\right)  -\left(
\begin{array}
[c]{cc}%
\A^{-1} & \K^{(k)}\\
\K^{(k)T} & \A^{-1}%
\end{array}
\right)  \right)  \mathcal{G}^{(k)T}
\end{equation*}
or
$
\mathcal{E}^{(k+1)}=\mathcal{G}^{(k)}\mathcal{E}^{(k)}\mathcal{G}^{(k)T}
$
for $k\ge 0$, where
$
\mathcal{E}^{(k)}  =\Var\left(  \f{Y}^{(k)}\right)  -\left(
\begin{array}
[c]{cc}%
\A^{-1} & \K^{(k)}\\
\K^{(k)T} & \A^{-1}%
\end{array}
\right).
$
Hence, by recursion,
$\mathcal{E}^{(k)}=\left(  \prod_{l=0}^{k-1}\mathcal{G}^{(l)}\right)
\mathcal{E}^{(0)}\left(  \prod_{l=0}^{k-1}\mathcal{G}^{(l)}\right)  ^{\text{T}}.%
$

Denote the polynomial of the block matrix by
    $\mathcal{P}^{(k+1)}=\left(  \prod_{l=0}^{k}\mathcal{G}^{(l)}\right)$
that satisfies%
\begin{align*}
\mathcal{P}^{(k+1)}  &  =\mathcal{G}^{(k)}\mathcal{P}^{(k)}
  =\left(
\begin{array}
[c]{cc}%
\alpha_{k}\G^{(k)} & \left(  1-\alpha_{k}\right)  \f{I}\\
\f{I} & \f{0}
\end{array}
\right)  \left(
\begin{array}
[c]{cc}%
\mathcal{P}_{11}^{(k)} & \mathcal{P}_{12}^{(k)}\\
\mathcal{P}_{21}^{(k)} & \mathcal{P}_{22}^{(k)}%
\end{array}
\right) 
\end{align*}
with
$\mathcal{P}^{(1)}  =\mathcal{G}^{(0)}=\left(
\begin{array}
[c]{cc}%
\G^{(0)} & \f{0}\\
\f{I} & \f{0}
\end{array}
\right).$
Thus 
\begin{align*}
\mathcal{P}_{11}^{(k+1)}  &  =\alpha_{k}\G^{(k)}\mathcal{P}_{11}^{(k)}+\left(
1-\alpha_{k}\right)  \mathcal{P}_{11}^{(k-1)}
  =\alpha_{k}\left(  \f{I}-\tau_{k}\M^{-1}\A\right)  \mathcal{P}_{11}^{(k)}+\left(
1-\alpha_{k}\right)  \mathcal{P}_{11}^{(k-1)}%
\end{align*}
with $\mathcal{P}_{11}^{(1)}=\G^{(0)}$, which shows that $\mathcal{P}_{11}^{(k+1)}=P_{k+1}$ by \eqref{RecursionErrCheby}.  Furthermore, this shows that the error in variance at the $k^{\text{th}}$ iteration has the specified form and convergence factor.

\bibliography{JRSSBcite_2014-10-22}
\bibliographystyle{chicago}

\end{document}